\documentclass[]{aastex631}

\usepackage{amsmath}
\usepackage{subfigure}
\usepackage{natbib}
\usepackage{todonotes}
\usepackage{booktabs}

\newcommand{\inspiralesigmahm}{\texttt{InspiralESIGMAHM}}
\newcommand{\inspiralesigma}{\texttt{InspiralESIGMA}}
\newcommand{\esigmahm}{\texttt{ESIGMAHM}}

\usepackage{hyperref}
\hypersetup{
    colorlinks=true,
    linkcolor=blue,
    filecolor=magenta,      
    urlcolor=blue,
    pdftitle={Overleaf Example},
    pdfpagemode=FullScreen,
    }
\usepackage{dcolumn}
\usepackage{bm}
\usepackage{orcidlink}
\usepackage{graphicx}

\begin{document}

\title[Eccentricity of Supermassive Black Hole Binaries]{An Accurate Modeling of Nano-hertz Gravitational Wave Signal from Eccentric Supermassive Binary Black Holes: An Essential Step Toward a Robust Discovery}

\author[0009-0005-9881-1788]{Mohit Raj Sah}
\affiliation{Department of Astronomy and Astrophysics, Tata Institute of Fundamental Research, Mumbai 400005, India}
\email{mohit.sah@tifr.res.in}

\author[0009-0006-9399-9168]{Akash Maurya}
\affiliation{International Centre for Theoretical Sciences, Tata Institute of Fundamental Research, Bangalore 560089, India}
\email{akash.maurya@icts.res.in}

\author[0000-0002-3373-5236]{Suvodip Mukherjee}
\affiliation{Department of Astronomy and Astrophysics, Tata Institute of Fundamental Research, Mumbai 400005, India}
\email{suvodip.mukherjee@tifr.res.in}

\author[0000-0001-5523-4603]{Prayush Kumar}
\affiliation{International Centre for Theoretical Sciences, Tata Institute of Fundamental Research, Bangalore 560089, India}
\email{prayush@icts.res.in}

\author[0009-0000-7559-7962]{Vida Saeedzadeh}
\affiliation{Department of Physics $\&$ Astronomy, Johns Hopkins University, Baltimore, MD 21218, USA}

\author[0000-0003-1746-9529]{Arif Babul}
\affiliation{Department of Physics and Astronomy, University of Victoria, 3800 Finnerty Road, Victoria, BC, V8P 1A1, Canada}
\affiliation{Infosys Visiting Chair Professor, Indian Institute of Science, Bangalore 560012, India}

\author[0000-0002-8115-8728]{Chandra Kant Mishra}
\affiliation{Department of Physics, Indian Institute of Technology Madras, Chennai 600036, India}
\affiliation{Centre for Strings, Gravitation and Cosmology, Department of
Physics, Indian Institute of Technology Madras, Chennai 600036, India}

\author[0000-0002-8406-6503]{Kaushik Paul}
\affiliation{Department of Physics, Indian Institute of Technology Madras, Chennai 600036, India}
\affiliation{Centre for Strings, Gravitation and Cosmology, Department of
Physics, Indian Institute of Technology Madras, Chennai 600036, India}

\author[0000-0001-5510-2803]{Thomas R. Quinn}
\affiliation{Astronomy Department, University of Washington, Box 351580, Seattle, WA, 98195-1580, USA}

\author[0000-0002-4353-0306]{Michael Tremmel}
\affiliation{School of Physics, University College Cork, College Road, Cork T12 K8AF, Ireland}

\begin{abstract}
The stochastic gravitational wave background (SGWB) in the nanohertz (nHz) regime, detectable by pulsar timing arrays (PTAs), provides a promising probe of the cosmic population of supermassive black hole binaries (SMBHBs). These binaries are expected to retain significant eccentricity throughout their evolution. We present a new technique to model the nHz SGWB by incorporating eccentricity into a multi-scale adaptive simulation framework. Using the time-domain eccentric waveform model \esigmahm{}, we generate realistic GW signals from astrophysical populations of SMBHBs. Unlike circular binaries, eccentric systems emit across multiple frequencies, introducing spectral correlations between frequency bins. These correlations provide a novel observational signature of the eccentricity distribution of the SMBHB population. In this work, we adopt simplified power-law models for the eccentricity distribution. While this does not capture the full complexity of galactic environments, it effectively highlights the key features of GW emission from eccentric binaries and their imprint on the SGWB. Our approach advances nHz GW signal modeling by incorporating eccentricity at small scales, enabling more realistic predictions and offering a new avenue for probing SMBHB astrophysics with future PTA observations.
\end{abstract}


\section{Introduction}

Supermassive black holes (SMBHs) are believed to reside at the centers of most, if not all, massive galaxies \citep{kormendy2013coevolution}. When these galaxies merge, the SMBHs at their centers form a binary system and over millions of years, under the influence of the galactic environment, these SMBH binaries (SMBHBs) inspiral and form tighter orbits \citep{volonteri2007evolution,volonteri2012formation,dosopoulou2017dynamical,ni2022astrid}. Once the component black holes (BHs) are sufficiently close, gravitational wave (GW) emission becomes the dominant mechanism driving their evolution \citep{kelley2017massive,burke2019astrophysics,chen2020dynamical}. When the binary separation shrinks below 0.1 parsec, they begin emitting GWs in the nanohertz (nHz) frequency range \citep{mannerkoski2019gravitational,burke2019astrophysics}. These nHz GWs are the target of the pulsar timing array (PTA) collaborations \citep{mclaughlin2013north,desvignes2016high,manchester2013parkes,joshi2018precision,xu2023searching,bailes2020meerkat}. Recently, several PTA collaborations reported evidence of nHz GW background (GWB) \citep{agazie2023nanograv,antoniadis2023second,zic2023parkes,xu2023searching}. The most probable source of this signal is believed to be the inspiraling population of the SMBHB \citep{sesana2008stochastic,sesana2013gravitational,burke2019astrophysics}. The astrophysical nHz GWB is expected to follow the underlying distribution of galaxies and exhibit spatially anisotropic signal \citep{sato2024exploring,sah2024imprints,semenzato2024cross}. This can be explored using multi-tracer correlated stacking \citep{Sah_2025} and cross-correlation with tracers of large scale structure such as Active Galactic Nuclei (AGN) and galaxies \citep{sah2024discovering,semenzato2024cross,sah2025route}.

 The inspiraling SMBHBs can be characterized by their component masses, their separation, their eccentricity, and their spins. Most existing analyses estimating the GW signal from SMBHBs neglect both eccentricity and spin in modelling the GW signal while performing nHz data analysis. These properties carry valuable information about the evolutionary history of the binary, since they are closely linked to the dynamics of galaxy mergers and interactions with the surrounding environment. While GW-driven evolution tends to circularize binary orbits, numerical simulations suggest that eccentricity tends to increase during evolution driven by stellar interactions \citep{chen2017efficient,gualandris2022eccentricity}.
Moreover, circumbinary disk interactions have been shown to be even more efficient at increasing the eccentricity and remain effective down to sub-parsec separations, where stellar scattering becomes less dominant \citep{siwek2024signatures}. This suggests that the circular approximation may not be an accurate representation of the SMBHB in the context of the GWs emitted by these binaries. The stochastic gravitational wave background (SGWB) generated by a population of eccentric binaries has been investigated in previous studies \citep{chen2017efficient,kelley2017gravitational,raidal2024eccentricity,falxa2025eccentric}. These studies employ analytical and semi-analytical approaches to estimate the SGWB signal. However, these estimates often fail to capture the full complexity of the GW signal from eccentric binaries. In particular, they typically overlook spectral distortions introduced by the finite frequency resolution resulting from the limited observation time of PTA experiments. In contrast, our approach employs a time-domain eccentric waveform model, \esigmahm{}, to generate realistic GW signals that more accurately reflect the spectral features of the SGWB generated by eccentric binaries.

An important consequence of this realistic modeling is the ability to capture the discrete nature of the SGWB spectrum. Studies suggest that the nHz SGWB is dominated by a few exceptionally bright sources \citep{sah2024imprints,semenzato2024cross}. Consequently, the SGWB is expected to exhibit discreteness in its spectrum. This arises from the fact that SMBHBs maintain highly stable orbits over the PTA observation period. In the case of circular orbits, these binaries act as nearly monochromatic GW sources. Due to the dominance of a few sources, significant Poisson fluctuations are expected in the SGWB density. In addition to spectral discreteness, eccentric binaries—unlike circular ones—introduce spectral correlations. This occurs because, for circular binaries, the power in each frequency bin is independent (as each binary emits monochromatically). In contrast, eccentric binaries emit GWs across multiple frequencies, coupling different bins in the spectrum.


 In this work, we propose a simulation-based framework to model the SGWB from eccentric SMBHBs across multiple physical scales. Our goal is to demonstrate a new technique for accurately modeling the SGWB signal from an eccentric population of SMBHBs, using a numerical relativity–based waveform model to capture the signal from eccentric binaries, which can be applied for analyzing nHz GW data. Importantly, this study does not dynamically evolve binaries from one orbit to another in the galactic environment. Rather, we construct an instantaneous distribution of sources. The number of sources at a given separation is treated as a free parameter and is assumed to be independent of eccentricity. As a result, our model does not capture the joint distribution of eccentricity and separation, as would be expected from dynamical evolution in realistic environments. However, incorporating such correlations is straightforward in our framework and can be implemented by drawing from a physically motivated distribution in future extensions. In this sense, our modeling marks a major improvement over previous analytical or semi-analytical estimates by enabling us to capture the more accurate multi-harmonic structure of GWs emitted by eccentric binaries. A key strength of this framework is its ability to predict not only the SGWB energy density spectrum but also the spectral correlations across frequency bins—an observable signature absent in circular binaries. We present results for different SMBHB populations with varying eccentricity distributions and analyze how the structure of the spectral covariance matrix encodes information about these distributions. These findings provide new insights into the impact of eccentricity on the SGWB and have important implications for PTA experiments aiming to detect and characterize the nHz SGWB.

 The paper is organized as follows: in Sec. \ref{Model and Sim}, we describe the simulation technique and the population models used in modeling the SGWB signal; in Sec. \ref{Obs}, we discuss the
results of the simulation; in Sec. \ref{prosp}, we discuss the prospect of detecting the eccentricity distribution of the SMBHB population. Finally, in Sec. \ref{Conc}, we discuss the conclusions and future prospects.

\section{Adaptive multi-scale technique \lowercase{{\large to}} calculate \lowercase{n}Hz GW SGWB signal \lowercase{{\large from}} simulations} \label{Model and Sim}

In this work, we follow the method proposed in \cite{sah2024imprints} to simulate the population of SMBHBs. We use a simulated galaxy catalog derived from large-scale cosmological simulations. These simulations, while covering large-scale structures (Gpc scale),  have relatively low resolution, due to computational limitations, and cannot resolve interactions below the galactic scale. To overcome this, we incorporate insights from high-resolution cosmological simulations with sub-kpc-scale gravitational softening, which accurately track the evolution, including the orbital dynamics, of galaxies and their SMBHs within smaller volumes.  Combining these two simulation regimes, we construct an adaptive multi-scale model: large-scale simulations provide the statistical distribution of host galaxies, while high-resolution simulations inform the detailed relationship between SMBHB properties and the host galaxy.

\begin{figure*}
    \centering
    \includegraphics[width=0.8\textwidth, height=0.35\textheight]{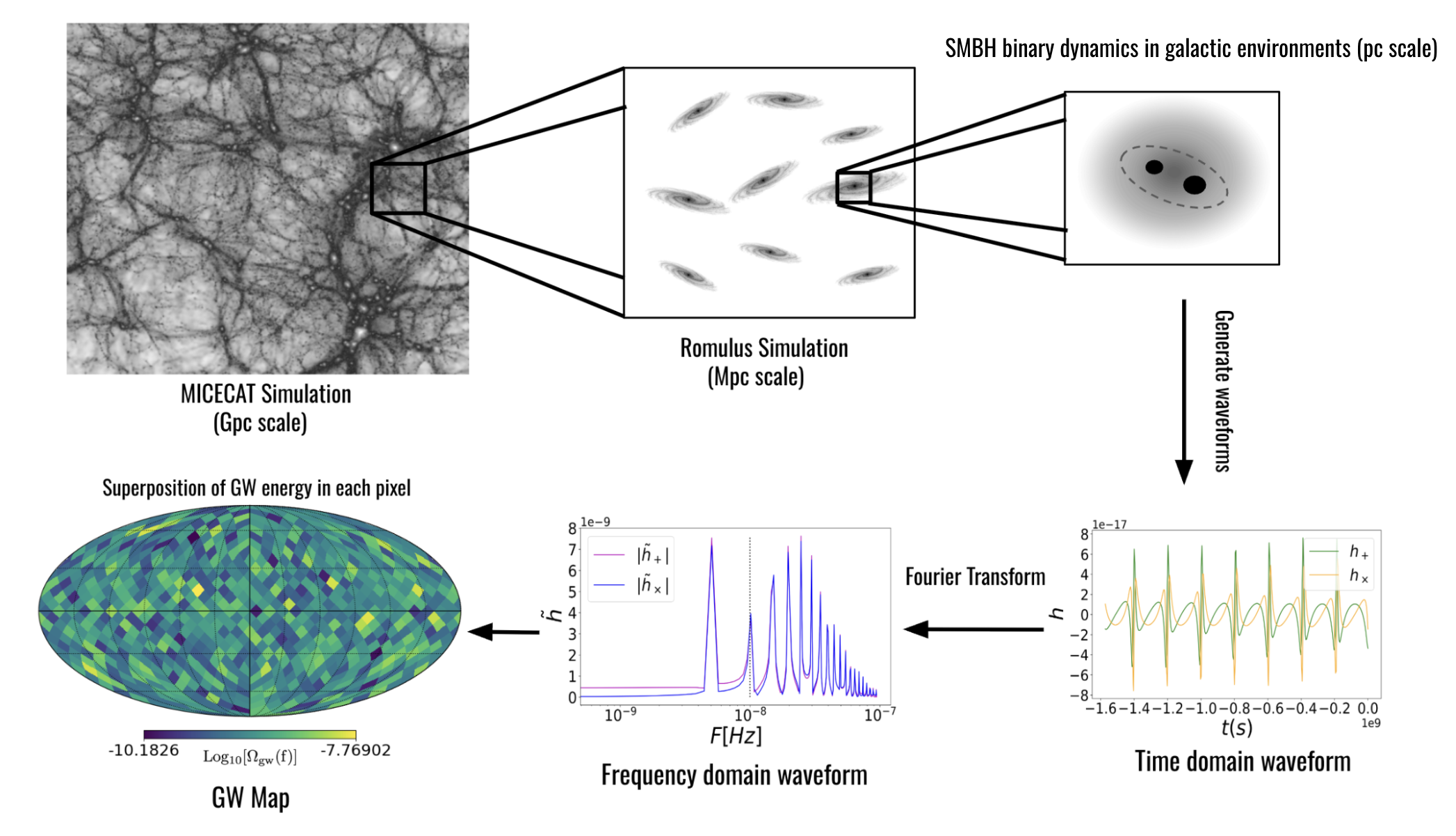}  
    \caption{Schematic diagram summarizing our simulation technique, showing the multi-scale framework linking Gpc-scale structure (\texttt{MICECAT}), Mpc-scale SMBH–galaxy co-evolution (\texttt{ROMULUS25}), and sub-pc binary dynamics including environment and eccentricity, to model the nHz SGWB. The SMBHB population is generated via a Monte Carlo sampling procedure based on the \texttt{MICECAT} simulated galaxy catalog. For each binary, a time-domain gravitational waveform is computed using eccentric waveform modeling. These waveforms are then truncated to match the chosen observation time and are subsequently Fourier transformed. Finally, the SGWB map is constructed by summing the GW energy density contributions from all sources within each sky pixel.}
    \label{fig:Diagram}
\end{figure*}

A schematic diagram summarizing this adaptive technique across scales, including the effects of eccentricity, is shown in Fig. \ref{fig:Diagram}. The steps involved in the multiscale modeling approach are as follows:
\begin{itemize}
    \item \textbf{Large-scale structure (Gpc scale):} We begin with the MICECAT catalog, derived from MICE-GC \citep{fosalba2015mice,crocce2015mice} dark matter simulations, to provide galaxy distributions and stellar mass functions over around 4 $\rm Gpc^{3}$ of comoving volume covering one-eighth of the sky.
    
    \item \textbf{Galaxy-scale physics (Mpc scale):} Using the \texttt{ROMULUS25} \citep{tremmel2017romulus,tremmel2020formation,jung2022massive,saeedzadeh2023cool,saeedzadeh2024shining} simulation, which is a high-resolution cosmological simulation that explores relatively small but more intricate scales, we can infer the connection between galaxy and SMBH properties. \texttt{ROMULUS25} extends over a uniform comoving volume of $\rm (25 ~ Mpc)^{3}$. The modeling of the SMBH mass and stellar mass of the galaxy is described in Sec. \ref{Model}.   

    \item \textbf{Environmental Effects (sub-pc scale) :} We can model the environmental effects (e.g., dynamical friction, stellar scattering, gas drag) affecting binary orbital decay and simulate distributions of binary separations and eccentricity. However, given the limitation, due to the resolution of the simulation, we do not directly simulate these processes. Instead, we approximate their influence by a simple parametric model introduced in \citet{saeedzadeh2024shining}.
    
    \item \textbf{Eccentricity of SMBHBs (sub-pc scale):} GWs emitted by eccentric binaries are modeled using a numerical relativity–based waveform model called \esigmahm{}. The details of the \esigmahm{} waveform modeling are provided in Sec.~\ref{Waveform}.
\end{itemize}

These components are integrated into an adaptive multi-scale framework to simulate the nHz SGWB signal with both spatial and spectral resolution.

\subsection{Modeling SMBHB Population} \label{Model}

SGWB is the superposition of all the GW sources that are unresolved as an individual source. The SGWB is defined as the GW energy density per unit logarithmic frequency divided by the critical energy density of the Universe.  The SGWB density can be written as a function of frequency $f$ and sky position $\hat n$ as \citep{phinney2001practical,sesana2008stochastic,christensen2018stochastic}
\begin{equation}
    \begin{aligned}
    \overline{\Omega}_{\rm GW}(f, \hat n) = \frac{1}{ \rho_c c^2} & \int \prod_{i}^{n} d\theta_i \int_{z_{\rm min}}^{\infty} dz \frac{d^2V}{dzd\omega} \left[\frac{1}{1+z} R_{\rm GW}(\Theta_n,z,\hat{n})\right] \left[\frac{1+z}{4 \pi d_{L}^{2} c}\right] \left[ f_r \frac{dE_{\rm{gw}}(\Theta_n,z, \hat n)}{df_r} \right],
    \end{aligned}
    \label{SGWB}
\end{equation}
where $R_{\rm GW}(\Theta_n,z,\hat{n})$ is the merger rate of the SMBHB with source properties $\Theta_n$, at redshift z, in the sky direction $\hat n$, $\omega$ represents the solid angle, $\frac{ dE_{\rm{gw}}(\Theta_n,z,\hat{n})}{df_r}$ is the GW energy emitted by the source per unit source frame frequency ($f_r = (1+z) f$), $\Theta_n$ = $\{\theta_{i}\}_{i=1}^{n}$ denotes the set of n GW source parameters, and $\rho_c c^{2} $ is the critical density of the universe. 

For a population of a stable, inspiraling SMBHB observed over a duration $T_{\rm obs}$, the equation can be rewritten as 

\begin{equation}
    \begin{aligned}
    \overline{\Omega}_{\rm GW}(f, \hat n) = &\frac{1}{\rho_c c^2}  \int \prod_{i}^{n} d\theta_i \int_{z_{\rm min}}^{\infty} dz \frac{d^2V}{dzd\omega} \frac{1}{c ~ T_{\rm obs}} \left[ \frac{d^{n+3}N(\Theta_n,z,\hat{n})}{d\Theta_n dV}\right] \times \left[f \frac{d\mathcal{E}_{\rm{gw}}(\Theta_n,z,\hat{n})}{df} \right],
    \end{aligned}
    \label{SGWB}
\end{equation}
where $\frac{ d^{n+3}N(\Theta_n,z,\hat n)}{ d\Theta_n dV}$ is the number of the binaries per unit comoving volume at redshift z, in the direction $\hat n$, and per unit source properties $\Theta_n$. The quantity $\frac{ d\mathcal{E}_{\rm{gw}}(\Theta_n,z,\hat n)}{df}$ is defined as 
\begin{equation}
    \frac{ d\mathcal{E}_{\rm{gw}}(\Theta_n,z,\hat n)}{df} = \frac{\pi c^{3}}{2 G} f^{2} (|\tilde{h}_{+}(f,\Theta_n,\hat n)|^{2} + |\tilde{h}_{\times}(f,\Theta_n,\hat n)|^{2} ),
\end{equation}
where $\tilde{h}_{+}(f,\Theta_n,\hat{n})$ and $\tilde{h}_{\times}(f,\Theta_n,\hat{n})$ are the Fourier transforms of the time-domain strain components for the two GW polarizations, computed over the observation time $T_{\rm{obs}}$. Details of the simulations and waveform modeling are provided in the next subsection. The term $\frac{dN^{n+3}(z, \Theta_n, \hat{n})}{d\Theta_n dV}$ can be modeled based on the relationship between SMBHBs and their host galaxy properties. It can be expressed as
\begin{equation}
    \begin{aligned}
        \frac{d^{6}N}{dM_{\rm BH} dq dV da} \propto & ~ \int dM_{*}  \frac{d^4N}{dM_* dV}
        P(M_{\rm BH}, q, \mathtt{a}, e | M_*, z)\\
        \propto & \int dM_{*}  \frac{d^4N}{dM_* dV}  P(M_{\rm BH} | M_*, z) P(q | M_*, z) P(\mathtt{a} | M_*, z) P(e | M_*, z),
        \label{eq:binary_population}
    \end{aligned}
\end{equation}
here, $P(M_{\rm BH} | M_*, z)$ describes the probability distribution of primary BH masses conditioned on the stellar mass of the host galaxy ($M_*$) and redshift ($z$). Similarly, $P(q | M_*, z)$ characterizes the distribution of binary mass ratios, while $P(\mathtt{a} | M_*, z)$ and $P(e | M_*, z)$ represent the distributions of the semi-major axis ($\mathtt{a}$) and the eccentricity ($e$), respectively. In the above equation, we assume that the parameters $M_{\rm BH}, q, \mathtt{a}$ and $e$ are uncorrelated. We parametrize $P(M_{\rm BH} | M_*, z)$ and $P(q | M_*, z)$ as follows:

\begin{equation}
    P( M_{\rm BH}|  M_{*},z) \propto \mathcal{N}\Big(\mathrm{Log}_{10}\Big(\frac{M_{\rm BH}}{M_{\odot}}\Big)| \mathrm{Log}_{10}\Big(\frac{M_{\mu}(M_*, z)}{M_{\odot}}\Big),\sigma_m \Big),
    \label{M1}
\end{equation}

\begin{equation}
     P(q|  M_{*},z) \propto \bigg\{
    \begin{array}{cl}
    & 1/q, \quad  0.01 < q < 1,\\
    & 0, ~~ \rm else ,
    \end{array}
    \label{q}
\end{equation}
where $\mathcal{N}$ denotes a normal distribution with mean $\mathrm{Log}_{10}\Big(\frac{M_{\mu}(M_*, z)}{M_{\odot}}\Big)$ and standard deviation $\sigma_m$. The mean BH mass is modeled as 

\begin{equation}    \mathrm{Log}_{10}\Big(\frac{M_{\mu}(M_*, z)}{M_{\odot}}\Big) = \eta + \rho~ \mathrm{Log}_{10}(M_*/10^{11} M_\odot). 
    \label{Mmu}
\end{equation}
The parameters $\eta$, and $\rho$ define the scaling relation between BH mass and the stellar mass of the galaxy. Similarly, the semi-major axis and the eccentricity distributions are modeled as follows.

\begin{equation}
    P(\mathtt{a}|  M_{*},z) \propto \mathtt{a}^{\delta},
    \label{semi-major}
\end{equation}
\begin{equation}
    P(e|  M_{*},z) \propto e^{\zeta},
    \label{ecc}
\end{equation}
where $\delta$ and $\zeta$ are the power-law indices of the distributions of semi-major axis and eccentricity, respectively. The distribution of semi-major axes is influenced by GW emission and several environmental processes—including stellar scattering, gas interactions, and dynamical friction—that govern the orbital hardening and evolution of binaries. Although a simple power-law may not accurately represent the true distribution of binary separation, for this analysis, this serves as a reasonable approximation because we presently do not aim to precisely model the effects of the environment. The adaptive multi-scale technique proposed here can be easily extended to an astrophysical scenario with eccentricity evolution with binary separation.  In Appendix \ref{sec:ecc_evol}, we present results from a scenario in which binaries evolve solely under GW emission, with eccentricity and separation assigned according to this evolution. More accurate treatments of environmental effects, such as those used in \cite{sampson2015constraining} and \cite{saeedzadeh2024shining}, can easily be included in this framework.

We perform Monte Carlo simulations of source properties, using the \texttt{MICECAT} galaxy catalog~\citep{crocce2015mice}. \texttt{MICECAT} is a simulated galaxy catalog that covers one-eighth of the sky and extends up to a redshift of 1.4. In addition to the source properties described above, we also sample the binary orbital inclination angle and the mean anomaly from uniform distributions over the ranges (0,$\pi$) and (0,$2\pi$), respectively. A subset of galaxies is randomly selected from the catalog. We select only those galaxies whose stellar mass exceeds that of the galaxy corresponding to a mean BH mass (given by Eq.~\eqref{Mmu}) greater than $10^{8} M_{\odot}$, since lighter BHs contribute insignificantly to the signal. For each selected galaxy, we assign an SMBHB by sampling its physical properties using Eqs.~\ref{M1}, \ref{q}, \ref{semi-major}, and~\ref{ecc}. Once the sources are sampled, we generate the time-domain waveform using the \esigmahm{} \citep{paul2024esigmahm} waveform model (details provided in the next subsection). The waveform is then sliced to match the desired observation time and Fourier transformed to obtain its frequency-domain signal. The SGWB map, $\Omega_{\rm GW}(f, \hat{n})$, is then constructed by summing the GW energy emitted by individual sources within each sky pixel.

\subsection{Waveform modeling of eccentric binary} \label{Waveform}

Due to the expectation that GW radiation reaction efficiently circularizes the orbits of BH binaries \citep{PhysRev.136.B1224} by the time they reach the sensitivity bands of the present ground-based detectors, the analyses of stellar mass BH binaries by the LIGO-Virgo-KAGRA (LVK) collaboration have primarily used only quasi-circular GW waveform models \citep{LIGOScientific:2018mvr, LIGOScientific:2020ibl, KAGRA:2021vkt}. However, significant efforts have been made over the past few years to include the effect of orbital eccentricity in GW waveform models ~\citep{Konigsdorffer:2006zt,Klein:2010ti,Huerta:2014eca,Moore:2016qxz,Hinderer:2017jcs,Hinder:2017sxy,Huerta:2017kez,Cao:2017ndf,  Moore:2018kvz,Tiwari:2020hsu,Chen:2020lzc,Yun:2021jnh, Albanesi:2023bgi,Carullo:2024smg}. Previous studies \citep{Huerta:2014eca, Mishra:2015bqa, Tanay:2016zog, Moore:2019xkm, Klein:2021jtd, Paul:2022xfy, Henry:2023tka,Sridhar:2024zms} developed inspiral-only eccentric models using the post-Newtonian (PN) theory, with \cite{Klein:2018ybm, Klein:2021jtd} including the effect of spin-precession as well.  A few eccentric, aligned-spin, inspiral-merger-ringdown (IMR) models have also been developed ~\citep{Huerta:2016rwp,Huerta:2017kez,Hinder:2017sxy,Chen:2020lzc,Chiaramello:2020ehz,Ramos-Buades:2021adz, Albertini:2023aol, paul2024esigmahm, Gamboa:2024hli, Planas:2025feq}. Assuming quasi-circularity near the merger, these include hybrid models with an eccentric inspiral piece attached to a quasi-circular merger-ringdown piece \citep{Huerta:2016rwp, Huerta:2017kez, Hinder:2017sxy, Chattaraj:2022tay, Manna:2024ycx, paul2024esigmahm}, and effective one body (EOB) models \citep{Chiaramello:2020ehz, Nagar:2021gss, Gamba:2024cvy, Nagar:2024dzj, Ramos-Buades:2021adz, Gamboa:2024hli, Liu:2019jpg, Liu:2021pkr, Liu:2023dgl}. \cite{Islam:2021mha} also developed a numerical relativity (NR) surrogate model built over equal-mass, aligned-spin, eccentric NR simulations. Additionally, there also exist phenomenological prescriptions to transform quasi-circular waveform models into eccentric models \citep{Setyawati:2021gom, Wang:2023ueg, Islam:2024bza, Islam:2024zqo}.

\esigmahm{} \citep{paul2024esigmahm} is an eccentric, aligned-spin, IMR waveform model with higher-order GW modes. The inspiral piece of the model, called \inspiralesigmahm{}, is described by a combination of post-Newtonian (PN) theory and self-force theory. Assuming moderate starting eccentricities that will decay by the late inspiral, this inspiral piece is smoothly attached with the quasi-circular plunge-merger-ringdown piece coming from the numerical relativity surrogate model \texttt{NRSur7dq4} \citep{Varma:2019csw} to produce the full IMR waveform model. The model contains $(2, \pm 1)$, $(3, \pm 3)$, $(3, \pm2)$, $(4, \pm 4)$ and $(4, \pm 3)$ GW modes in addition to the dominant $(2, \pm 2)$ modes. Additionally, \esigmahm{} agrees well with many aligned-spin quasi-circular and eccentric SXS NR simulations \citep{Boyle:2019kee}.

\inspiralesigmahm{} uses state-of-the-art results from PN theory and self-force theory. It numerically solves for the orbital dynamics of the binary, computing the relative separation ($r$), angular position ($\phi$), and their time derivatives $\dot{r}$ and $\dot{\phi}$ as a function of time. These are then used to compute the GW strain by using the general-orbit PN expressions of the spin-weighted spherical harmonic modes $h_{\ell m}$ that are written in terms of dynamical variables such as $r, \dot{r}, \phi$, and $\dot{\phi}$.
The orbital evolution in the model is parameterized using the generalized quasi-Keplerian (QK) parameterization \citep{damour-1985} where the binary is described via the orbital variables like eccentricity ($e$), eccentric anomaly ($u$), mean anomaly ($l$),  and a PN parameter $x=(G M \Omega/c^3)^{2/3}$, where $M$ is the total mass, $\Omega$ is the GW half-frequency of the binary \citep{Blanchet:2023bwj}, $G$ is Newton's gravitational constant, and $c$ is the speed of light. The following equations give the \textit{conservative dynamics} (i.e., without taking into account the loss of energy and angular momentum via GW emission) of the binary,

\begin{align}
\label{eq:relative_separation}
\frac{r}{M}&= \frac{1-e\,\cos u}{x} + \sum_{i=1}^{3}a_{i}x^{i-1}\,, \\
\label{eq:Kepler_equation}
l &= u -e\, \sin u + \sum_{i=2}^{3} b_{i}x^{i}\,, \\
\label{eq:phi_dot}
M \dot{\phi}&= x^{3/2}\bigg[\sum_{i=0}^{3}c_{i}x^{i}+\mathcal{O}\Big(x^{4 }\Big)\bigg]\,, \\
\label{eq:l_dot}
M\dot{l}&=x^{3/2}\bigg[1+\sum_{i=1}^{3}d_{i}x^{i}+\mathcal{O}\Big(x^{4}\Big)\bigg]\,,
\end{align}
where the PN coefficients $a_i, b_i, c_i$ and $d_i$ are the PN corrections up to 3PN (\cite{Hinder:2008kv, Henry:2023tka}), and the index $i$ can take half-integer values. 

Assuming an adiabatic inspiral in which the radiation reaction time scales are much larger than the orbital time scales, the gradual loss of energy and angular momentum via GW emission yields the following orbit-averaged evolution equations for $x$ and $e$, describing the \textit{radiative dynamics} of the system, 

\begin{align}
   \label{eq:x_dot}
    M\dot{x} &= x^5\,\bigg[\sum_{i=0}^{4} y_{i}x^{i} +\mathcal{O}\Big(x^{9/2}\Big)\bigg]\,, \\
    \label{eq:e_dot}
    M\dot{e}&=x^{4} \bigg[\sum_{i=0}^{3}z_{i}x^{i} + \mathcal{O}\Big(x^4\Big)\bigg] \,.
\end{align} 
Here, $y_i$ and $z_i$ encode eccentric non-spinning and spinning corrections up to 3PN order \citep{Henry:2023tka}. Besides, $y_i$ also encodes the 3.5PN cubic-in-spin (SSS), 4PN non-spinning, spin-orbit (SO), and spin-spin (SS) terms for quasi-circular binaries \citep{Cho:2022syn, Blanchet:2023bwj, paul2024esigmahm}, and 4PN quasi-circular, non-spinning self-force corrections \citep{Fujita:2012cm, Huerta:2016rwp}.  Thus, given the starting GW frequency $(f_0 = \Omega_0/\pi)$, phase ($\phi_0$), eccentricity $(e_0)$ and mean anomaly $(l_0)$ as initial conditions, the inspiral dynamics of the binary can be obtained by numerically solving Eq. \eqref{eq:relative_separation}-\eqref{eq:e_dot}.

Finally, the GW polarizations $h_+$ and $h_\times$ are obtained from the superposition of the spin-weighted spherical harmonic modes $h_{\ell m}$ as,

\begin{equation}
    \label{eqn:SWSH}
    h_+(t; \Theta, \Phi) - i h_\times(t; \Theta, \Phi) = \sum_{\ell=2}^\infty \sum_{m=-\ell}^{\ell} h_{\ell m}(t) \, Y_{-2}^{\ell m}(\Theta, \Phi),
\end{equation}
where $Y_{-2}^{\ell m}$ are the $-2$ spin-weighted spherical harmonics \citep{NewmanPenrose}, which depend on the polar and the azimuthal angles $\Theta$ and $\Phi$ respectively that specify the binary's orientation with respect to the observer. In addition to the dominant $(\ell=2,m=\pm2)$ mode, \inspiralesigmahm{} also has $(2, \pm 1)$, $(3, \pm 3)$, $(3, \pm2)$, $(4, \pm 4)$ and $(4, \pm 3)$ modes. These modes have eccentric non-spinning and spinning instantaneous contributions \citep{Mishra:2015bqa,Paul:2022xfy, Henry:2023tka,paul2024esigmahm} and quasi-circular hereditary contributions \citep{Henry:2022dzx} till 3.5PN. The $(2,\pm 2)$ modes also contain the latest quasi-circular, non-spinning 4PN terms as reported in \cite{Blanchet:2023bwj, Blanchet:2023sbv}.

\begin{figure}
  \centering
  \subfigure[]{\label{fig:Wave0_a}
    \includegraphics[width=0.45\linewidth]{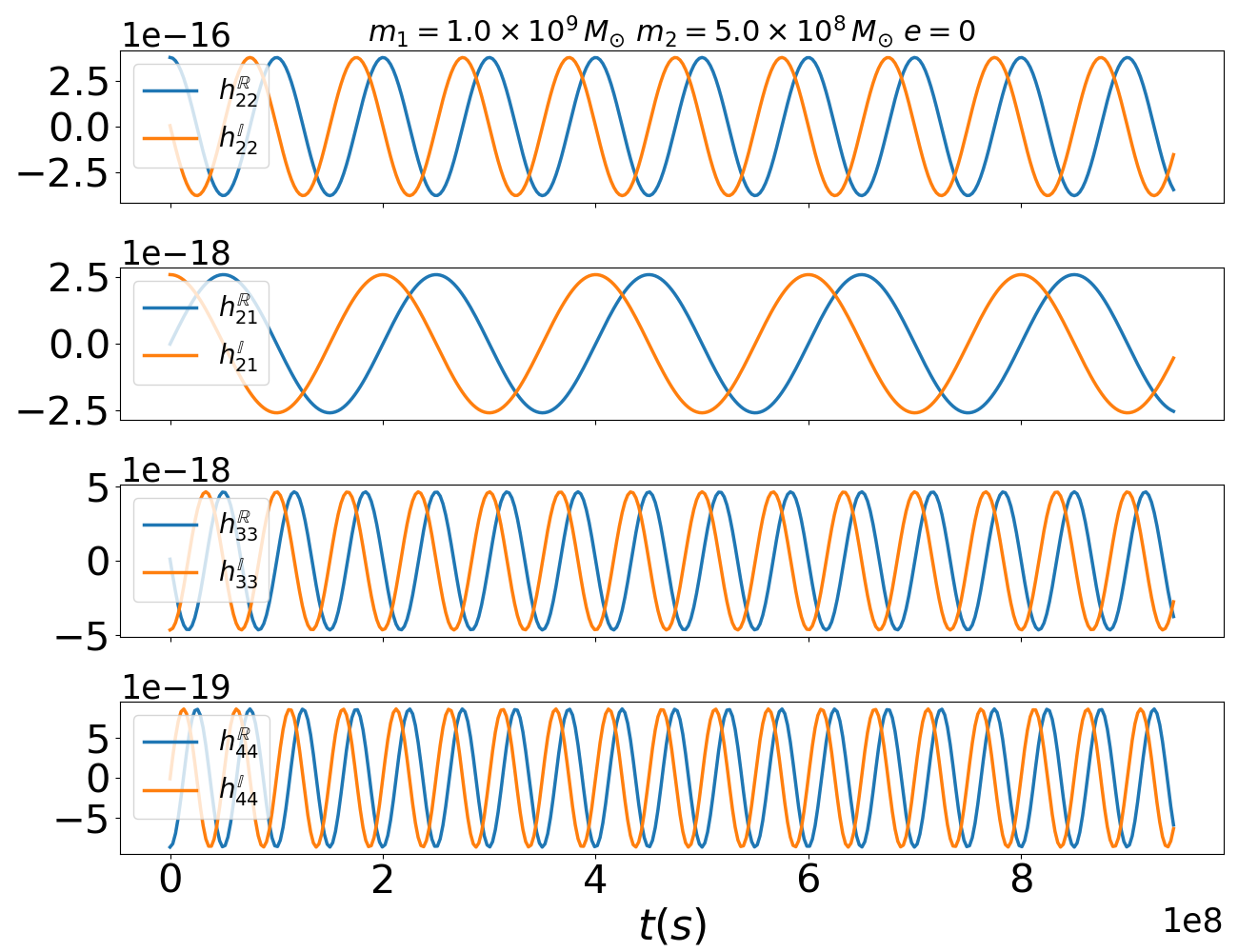}}
  \subfigure[]{\label{fig:Wave0_b}
    \includegraphics[width=0.45\linewidth]{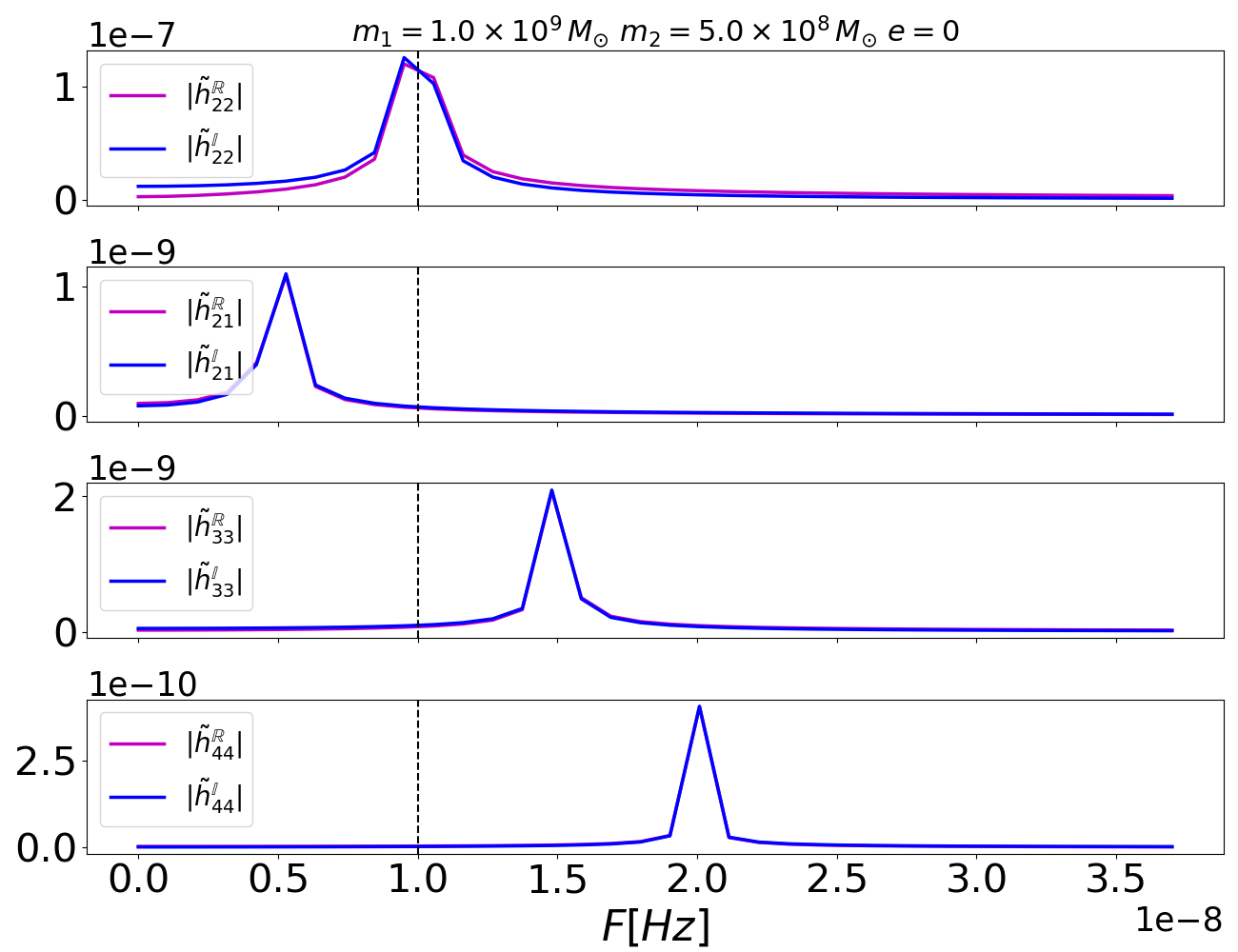}}
    \caption{Time-domain waveforms generated by \inspiralesigmahm{} along with their Fourier transforms for a circular binary ($e = 0$). $h^{\mathbb{R}}_{\ell m}$ and $h^{\mathbb{I}}_{\ell m}$ represent the real and imaginary parts of the $(\ell,m)$ GW mode, while $|\tilde{h}^{\mathbb{R}}_{\ell m}|$ and $|\tilde{h}^{\mathbb{I}}_{\ell m}|$ denote the absolute values of their respective Fourier transforms. The $(2,2)$ mode is shown along with the next three dominant modes: $(2,1)$, $(3,3)$, and $(4,4)$.}
  \label{fig:Wave0}
\end{figure}

\begin{figure}
  \centering
  \subfigure[]{\label{RA}
    \includegraphics[width=0.45\linewidth]{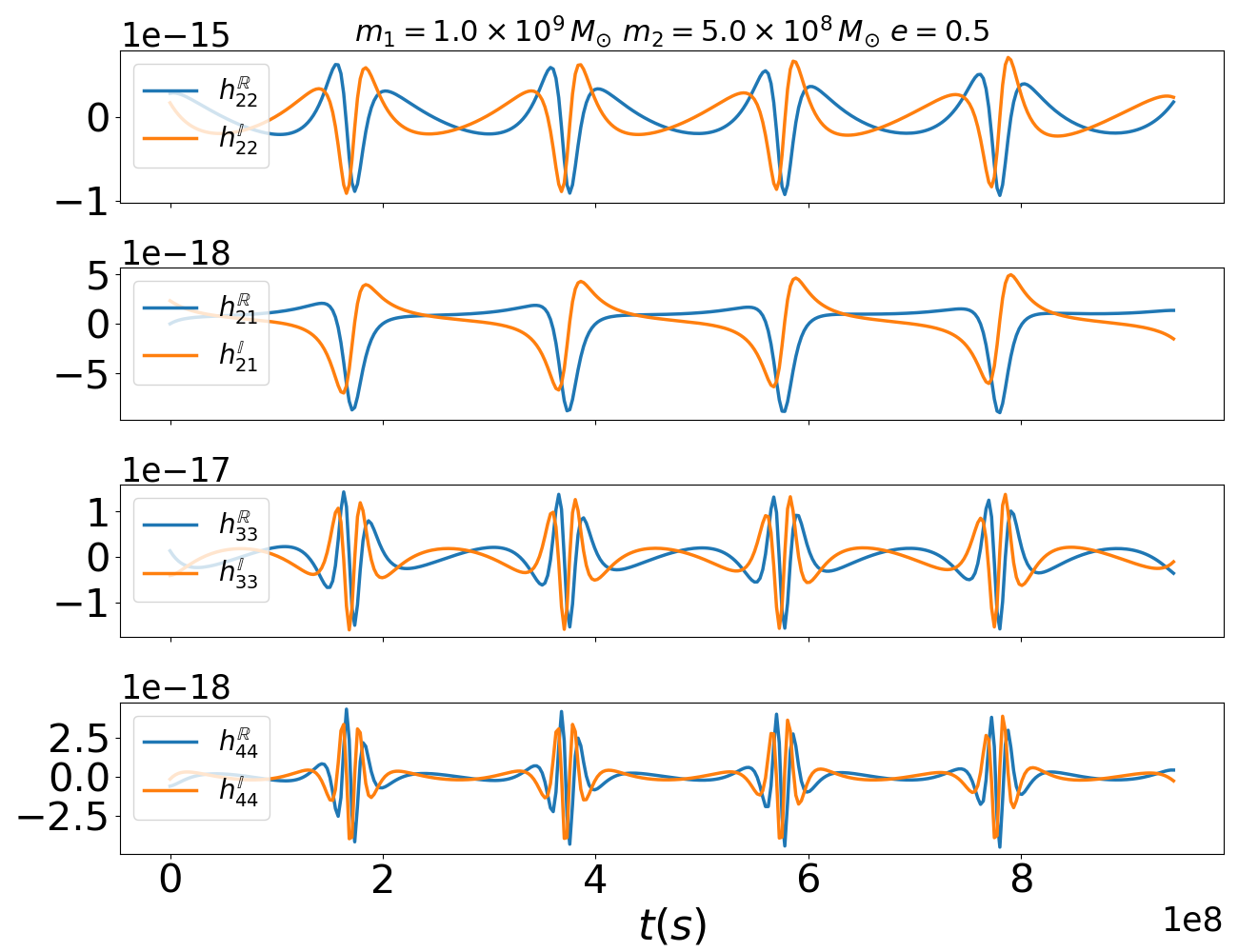}}
  \subfigure[]{\label{PA}
    \includegraphics[width=0.45\linewidth]{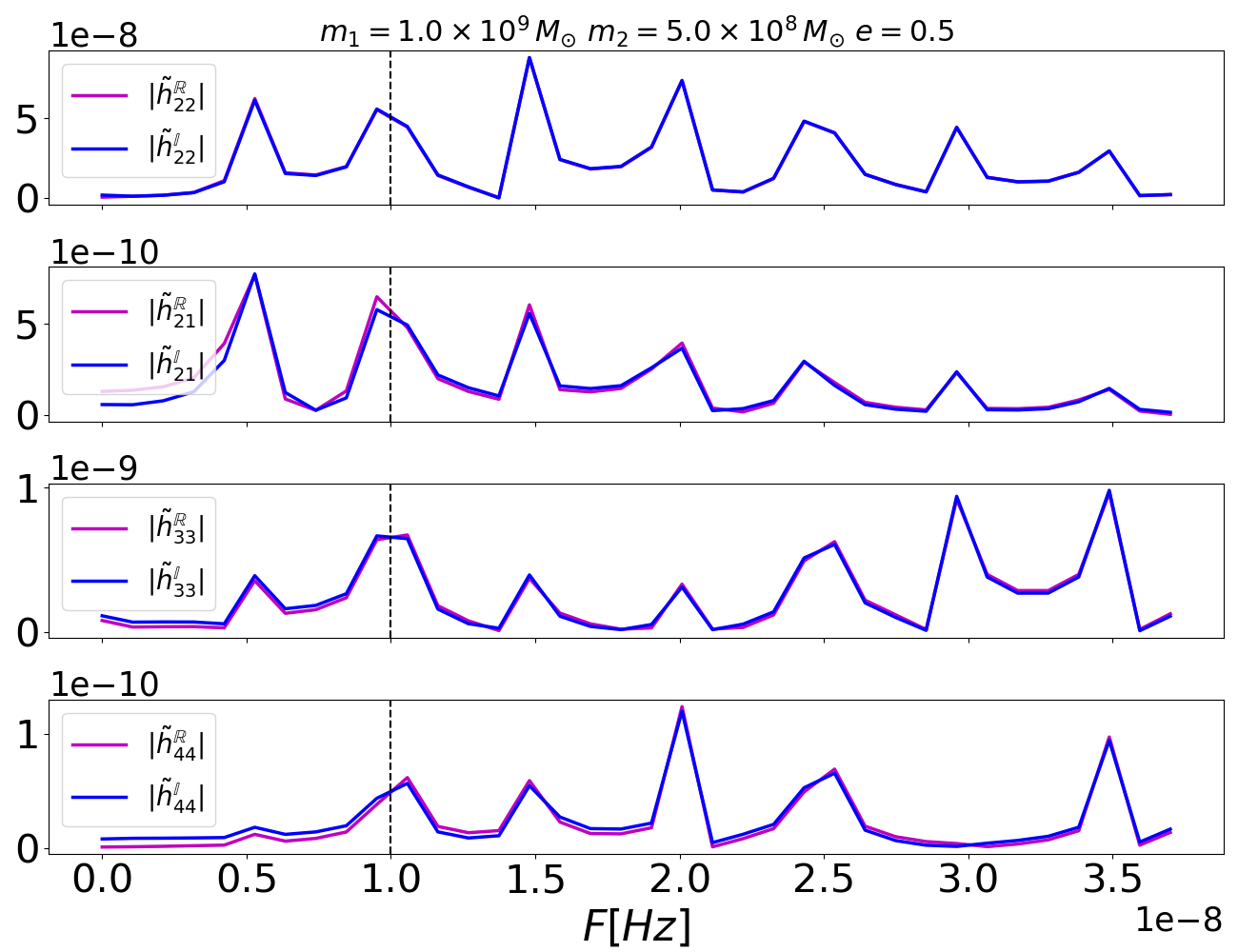}}
    \caption{Time-domain waveforms generated by \inspiralesigmahm{} along with their Fourier transforms for an eccentric binary ($e$ = 0.5). $h^{\mathbb{R}}_{\ell m}$ and $h^{\mathbb{I}}_{\ell m}$ represent the real and imaginary parts of the $(\ell,m)$ GW mode, while $|\tilde{h}^{\mathbb{R}}_{\ell m}|$ and $|\tilde{h}^{\mathbb{I}}_{\ell m}|$ denote the absolute values of their respective Fourier transforms. The $(2,2)$ mode is shown along with the next three dominant modes: $(2,1)$, $(3,3)$, and $(4,4)$.}
  \label{fig:Wave0.5}
\end{figure}

In Fig.~\ref{fig:Wave0} and Fig.~\ref{fig:Wave0.5}, we show the time-domain waveforms generated using \inspiralesigmahm{} and its Fourier transform for a system with total mass $1.5 \times 10^9 M_\odot$ and with a signal length of 30 years. In both the figures, we display the dominant $(2,2)$ mode along with the next three most significant modes: $(2,1)$, $(3,3)$, and $(4,4)$. Fig. \ref{fig:Wave0} illustrates the scenario where binary is in a circular orbit ($e = 0$). In Fig.\ref{fig:Wave0_a}, we show the real and imaginary parts of the $(\ell, m)$ GW modes, denoted by $h^{\mathbb{R}}_{\ell m}$ and $h^{\mathbb{I}}_{\ell m}$, respectively. Fig.\ref{fig:Wave0_b} shows the absolute values of the corresponding Fourier transforms, denoted as $|\tilde{h}^{\mathbb{R}}_{\ell m}|$ and $|\tilde{h}^{\mathbb{I}}_{\ell m}|$. The binary emits GWs that are nearly monochromatic, as clearly reflected in the narrow peaks of the Fourier-transformed waveforms. The small spread around the central GW frequency is due to the finite frequency resolution imposed by the limited duration of the signal.

The frequency resolution $(\Delta f)$, as well as the lowest frequency bin, is set by the inverse of the observation time: $\Delta f = 1/T_{\rm obs}$. In Fig.~\ref{fig:Wave0.5}, we show the GW modes for an eccentric orbit with $e = 0.5$, keeping all the other parameters unchanged. Unlike the circular case, the binary now emits over a broad range of frequencies. The higher-order modes are at least two orders of magnitude weaker than the $(2,2)$ mode in both scenarios. Hence, for computational efficiency, we use only the $(2,2)$ mode of \inspiralesigmahm{}, in the rest of the paper.

We also remark that we use the model to generate SMBHB waveforms with eccentricities as high as $e=0.8$.  However, it is not clear \textit{a priori} if the model can faithfully generate physically accurate waveforms until such high eccentricities given the resummed eccentricity expanded eccentric hereditary contributions in the radiative piece of the model \citep{Henry:2023tka}. Therefore, we also present a quantitative analysis of the accuracy of the model in Appendix \ref{appendix:accuracy_esigmahm} across the binary parameter space considered in this work. 

\begin{figure}
    \centering
    \includegraphics[width=11cm]{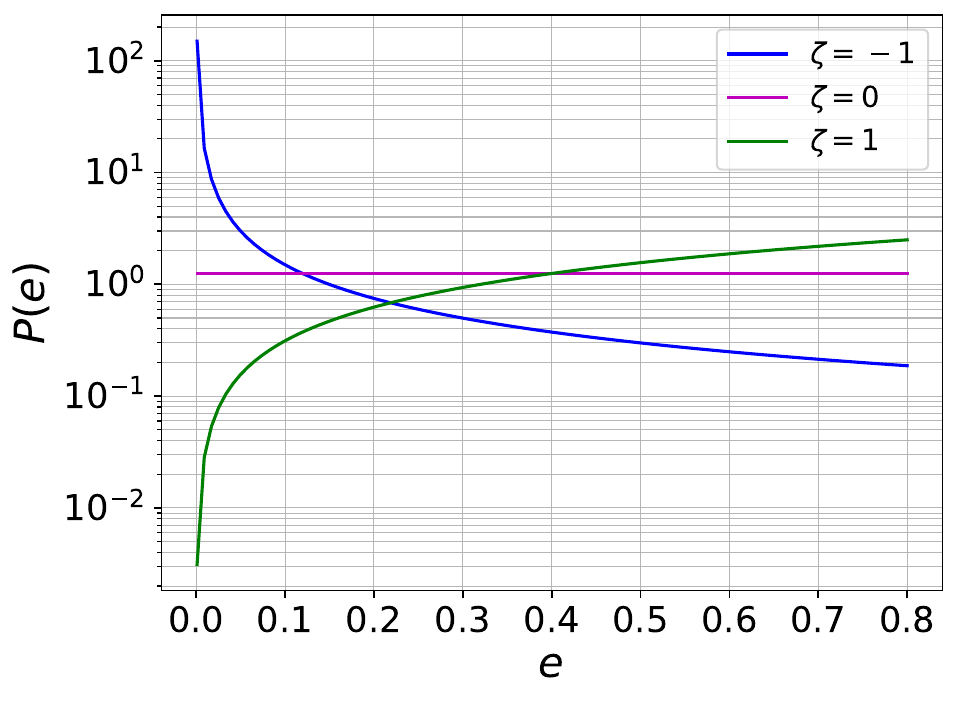}
    \caption{Eccentricity distributions of SMBHBs modeled using a power-law probability density function, $P(e) \propto e^{\zeta}$. Distributions are shown for three values of the power-law index: $\zeta = -1$ (favoring nearly circular binaries), $\zeta = 0$ (uniform distribution), and $\zeta = 1$ (favoring highly eccentric binaries).}
    \label{fig:Pe}
\end{figure}

\begin{figure}
    \centering
    \includegraphics[width=14cm]{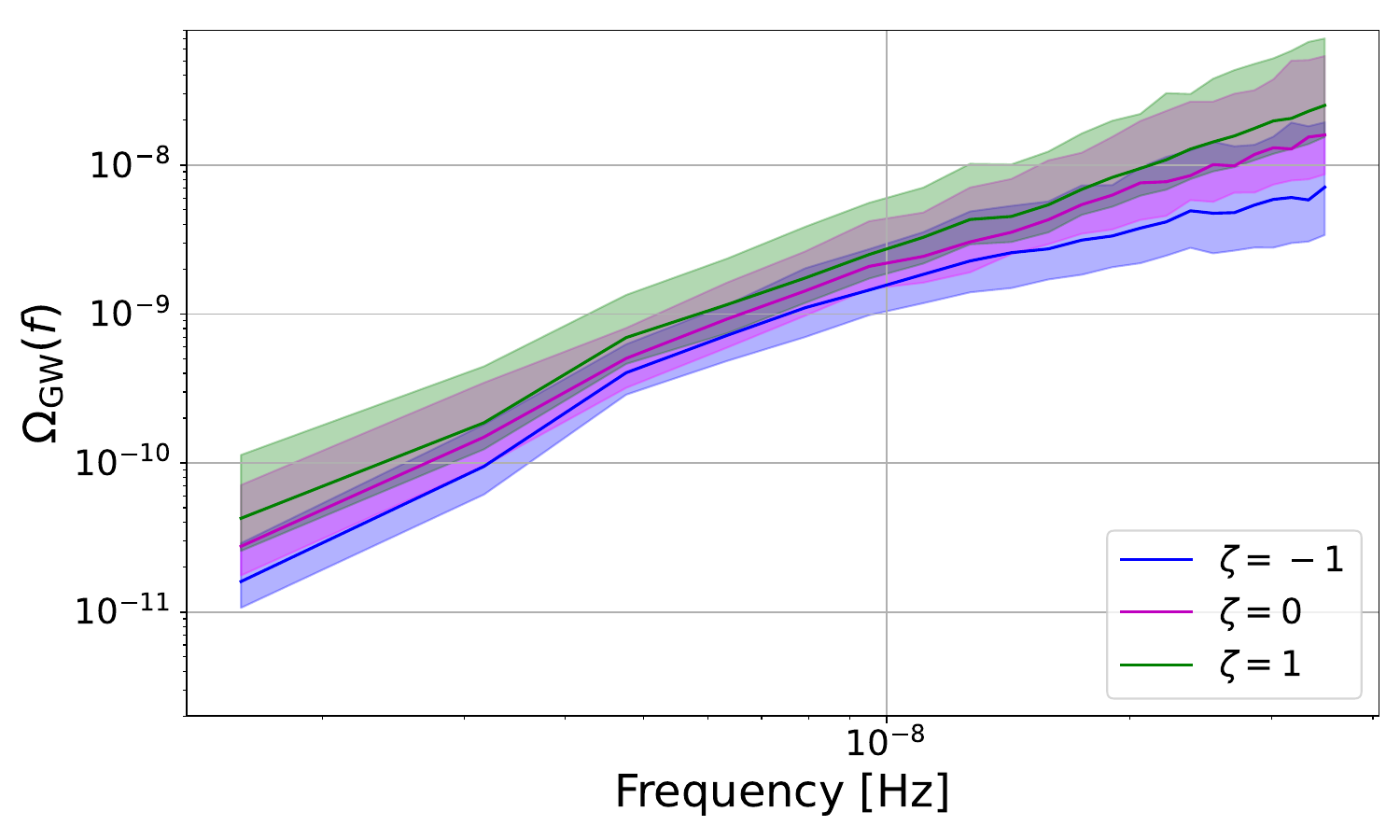}
    \caption{SGWB spectrum, $\Omega_{\rm GW}(f)$, for three different eccentricity distributions characterized by $\zeta = -1$, $\zeta = 0$, and $\zeta = 1$. The solid lines represent the median $\Omega_{\rm GW}(f)$ obtained from 300 Monte Carlo realizations, while the shaded regions denote the $68\%$ credible intervals.}
    \label{fig:Omega}
\end{figure}

\section{Observable Signatures \lowercase{{\large of}} Eccentricity \lowercase{{\large on}} \lowercase{{\large n}}-Hz SGWB signal} \label{Obs}
\subsection{SGWB Density from a population of Eccentric Binary}

In Fig. \ref{fig:Pe}, we illustrate the power-law distributions of eccentricity, $P(e) \propto e^{\zeta}$. We adopt this form because it is scale-free, making it well-suited for systems without a preferred eccentricity scale and allowing flexible modeling of various binary formation scenarios. We show $P(e)$ for three different values of $\zeta$: $\zeta = -1$, $0$, and $1$. The case with $\zeta = -1$ corresponds to a population dominated by nearly circular binaries, where low eccentricities are strongly favored. The case $\zeta = 0$ represents a uniform distribution in eccentricity, with all values equally likely within the allowed range. Finally, the $\zeta = 1$ case corresponds to a population dominated by highly eccentric binaries. We restrict the eccentricity in the range $0.001 \leq e \leq 0.8$; the lower bound is chosen to avoid divergence at $e = 0$ for negative values of $\zeta$, where the distribution becomes singular.

Fig. \ref{fig:Omega} shows the SGWB energy density spectrum, $\Omega_{\rm GW}(f)$, for three different eccentricity distributions characterized by $\zeta = -1$, $0$, and $1$.  Throughout this work, we adopt the following fiducial values for the population parameters: $\eta = 8$, $\rho = 1$, $\delta = 3.8$, and $\sigma_m = 0.4$. The values of $\eta$, $\rho$, and $\sigma_m$ are motivated by the various simulations and observational studies of the $M_{*}-M_{\rm BH}$ relation for both isolated and binary SMBHs \citep{reines2015relations,habouzit2021supermassive,saeedzadeh2023shining,kozhikkal2024mass}. The value $\delta = 3.8$ is chosen to reproduce the shape of the power-law spectrum that best fits the NANOGrav 15-yr data release \citep{agazie2023nanograv}. The spectrum is obtained from 300 Monte Carlo realizations, each with 25000 sources occupying the part of the sky covered by \texttt{MICECAT}. The choice of 25000 sources ensures that the mean SGWB spectrum is comparable to one-eighth (corresponding to the sky fraction covered by \texttt{MICECAT}) of the estimate from the NANOGrav 15-yr data release \citep{agazie2023nanograv}. The solid lines in the figure represent the median $\Omega_{\rm GW}(f)$ over 300 realizations, while the shaded bands denote the $68\%$ credible intervals. As evident from the figure, the overall amplitude of $\Omega_{\rm GW}(f)$ increases with increasing eccentricity. This trend differs from the scenarios where the eccentricity co-evolves with binary separation due to GW emission and environmental interactions; binaries tend to spend less time at larger separations. The faster orbital decay driven by high eccentricity reduces the population of widely separated binaries, which primarily emit at low frequencies. As a result, the SGWB power becomes increasingly suppressed at the lower frequencies with increasing eccentricity of the binary population. Moreover, the eccentricity itself evolves with binary separation. In contrast, the simplified example shown in this paper assumes an eccentricity distribution that is independent of binary separation and consequently does not capture this effect. To illustrate the implications of an evolutionary coupling between binary eccentricity and separation, Appendix. \ref{sec:ecc_evol} presents an example in which each binary is assigned an eccentricity and orbital separation based on the evolution of its orbit solely under GW emission.

Within this simplified framework, the value of $\zeta$ controls the eccentricity distribution. The $\zeta = 1$ case, representing a population dominated by highly eccentric binaries, yields the strongest signal across all frequencies, whereas the $\zeta = -1$ case, dominated by nearly circular binaries, exhibits the weakest signal. This enhancement in $\Omega_{\rm GW}(f)$ occurs because eccentric binaries emit GW radiation more efficiently than their circular counterparts.

In our future work, we will explore the dependency of the eccentricity and environmental effect in the evolution of the binaries, which may be useful to understand the degeneracy with spectral features arising from environmental interactions, such as gas dynamics and stellar hardening \citep{chen2024galaxy}. Both mechanisms accelerate binary evolution and redistribute GW power across frequencies, making it difficult to disentangle the contributions of eccentricity and environmental effects based solely on the isotropic SGWB spectrum. However, as we demonstrate in the subsequent sections, the spectral correlations induced by eccentric binaries provide a complementary observable that can help break this degeneracy.

\subsection{A new observable from eccentric SMBHBs: Spectral correlation of the anisotropic SGWB}

\begin{figure}
  \centering
  \subfigure[]{\label{fig:CovMinus1}
    \includegraphics[width=0.45\linewidth,trim={0.cm 0 0 0.cm},clip]{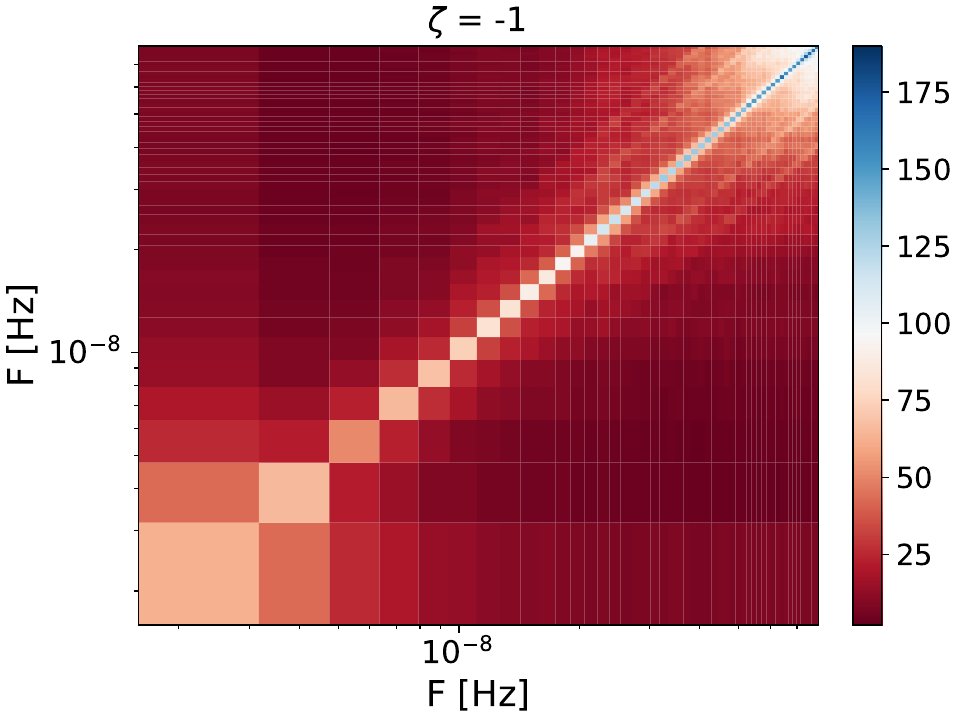}}
  \subfigure[]{\label{fig:Cov0}
    \includegraphics[width=0.45\linewidth,trim={0.cm 0cm 0cm 0.cm},clip]{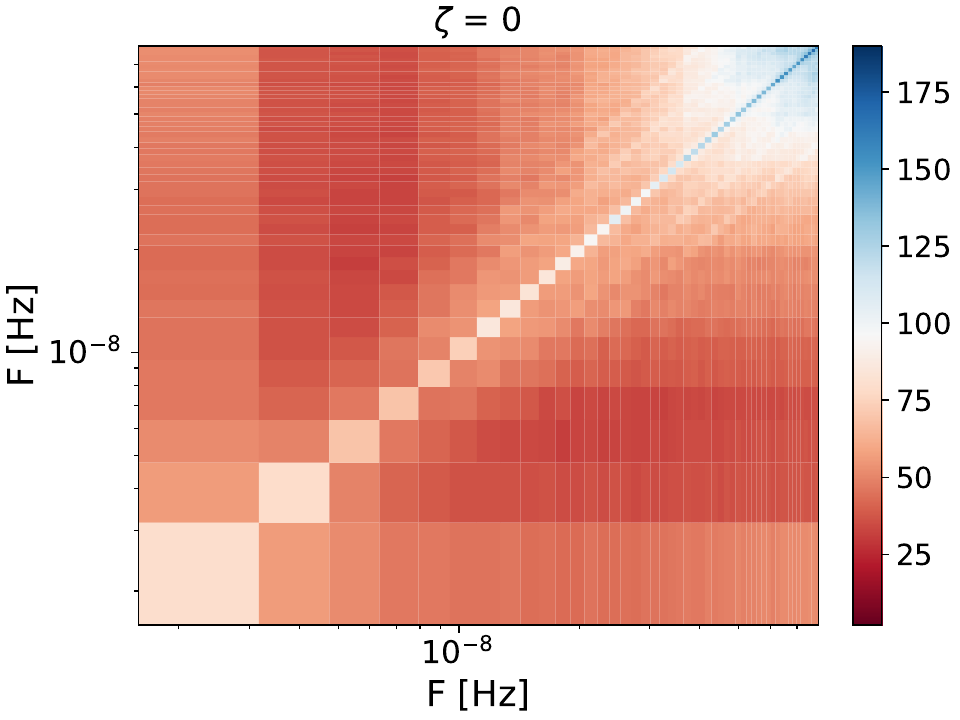}}\\
  \subfigure[]{\label{fig:Cov1}
    \includegraphics[width=0.45\linewidth,trim={0.cm 0cm 0cm 0.cm},clip]{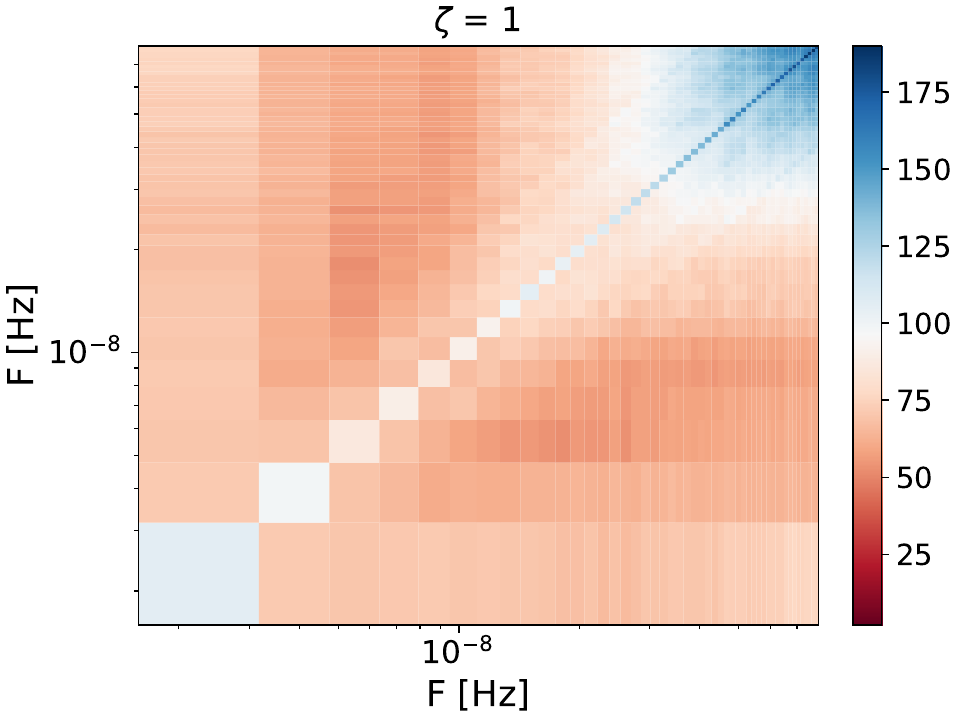}}
  \caption{Normalized spectral covariance matrix $C_{\rm N}(f_1, f_2)$ (defined in Eq. \ref{norm_cross}) for three different eccentricity distributions characterized by $\zeta = -1$, $0$, and $1$. The spectral correlations become increasingly prominent with increasing eccentricity, reflecting the broader harmonic content of GW signals from eccentric binaries.}
  \label{fig:Matrix}
\end{figure}

An SMBHB emits GWs across a broad range of frequencies throughout its inspiral, from formation to merger. For most SMBHB systems emitting in the PTA frequency band, the observational timescale—typically a few decades—captures only a few very stable orbits. If the binary is in a circular orbit, it emits GWs that are nearly monochromatic over short time intervals. In contrast, an eccentric binary emits GWs at multiple harmonics of its orbital frequency, leading to a multi-frequency signal even within a short observation window.

Since the number of binaries contributing to the SGWB is finite, their distribution varies across the sky, giving rise to fluctuations in the SGWB signal. For a population of binaries in purely circular orbits, each system emits monochromatically, and because these systems evolve independently, the SGWB signals at different frequencies are expected to be uncorrelated. However, for a population with non-zero orbital eccentricity, each binary contributes power to multiple frequencies simultaneously. As a result, the SGWB fluctuations at different frequencies become correlated.
This correlation can be quantified by the spectral covariance defined as follows
\begin{equation}
    C(f_1,f_2) = \frac{1}{N_{\rm pix}} \sum\limits_i^{N_{\rm pix}} \Big(\Omega^{i}_{\rm GW}(f_1) - \overline{\Omega}_{\rm GW}(f_1)\Big) \Big(\Omega^{i}_{\rm GW}(f_2) - \overline{\Omega}_{\rm GW}(f_2)\Big),
    \label{cross}
\end{equation}
where $N_{\rm pix}$ represents the number of pixels. We also define the normalized spectral covariance as

\begin{equation}
    C_{\rm N}(f_1,f_2) = \frac{C(f_1,f_2)}{\overline{\Omega}_{\rm GW}(f_1)~ \overline{\Omega}_{\rm GW}(f_2)},
    \label{norm_cross}
\end{equation}

where $\Omega^{i}_{\rm GW}(f)$ denotes the SGWB energy density spectrum in pixel $i$, obtained by summing the contributions from all sources within that pixel, and $\overline{\Omega}_{\rm GW}(f)$ represents the average SGWB energy density spectrum per pixel, computed by averaging over all sky pixels.

In Fig.~\ref{fig:Matrix}, we present the normalized spectral covariance matrix ($C_{\rm N}(f_1,f_2)$), which captures the covariance of the SGWB energy density. This is computed using the HEALPix framework \citep{gorski2005healpix,zonca2019healpy} with a resolution parameter of $N_{\rm side} = 16$, corresponding to $N_{\rm pix} = 3072$ pixels and a maximum multipole of $\ell_{\rm max} = 48$. Of these, approximately 384 pixels overlap with the sky coverage of the \texttt{MICECAT} catalog. This pixelation roughly corresponds to an array of $\sim$2000 pulsars, since the number of pulsars scales as $N_{\rm p} \sim \ell_{\rm max}^2$. This is within the reach of the Square Kilometre Array (SKA), which is expected to monitor thousands of millisecond pulsars with high timing precision \citep{smits2009pulsar}. The matrix is obtained by averaging over 300 independent Monte Carlo realizations of the source population, for three different eccentricity distributions. While varying the eccentricity distributions, all other properties of the sources are kept identical across the 300 realizations. We assume an observation time of $T_{\rm obs} = 20$ yr, consistent with projected SKA timelines \citep{lazio2013square}, corresponding to a frequency resolution of $\Delta f \approx 1.58 \times 10^{-9}$ Hz.

The structure of the spectral covariance matrix evolves significantly with the eccentricity distribution of the SMBHB population, parameterized by the power-law index $\zeta$. For $\zeta = -1$, which corresponds to a population dominated by nearly circular binaries ($P(e) \propto e^{-1}$), the covariance matrix is largely diagonal, indicating minimal spectral correlations. 
This is consistent with theoretical expectations given that circular binaries emit GWs at a single characteristic frequency, resulting in negligible spectral correlation. However, due to limited spectral resolution, the resulting spectrum of even circular binaries is not strictly monochromatic and exhibits small but non-negligible correlations across nearby frequency bins.

In contrast, a uniform eccentricity distribution ($\zeta = 0$) induces moderate off-diagonal correlations, reflecting the broader spectral correlation of the GW signal from eccentric binaries. These systems radiate at multiple frequencies, leading to spectral leakage across neighboring frequency bins. The effect becomes even more pronounced for $\zeta = 1$, where the binary population is highly eccentric, with a population of low eccentric binaries strongly suppressed. In this case, the covariance matrix exhibits strong off-diagonal features. The change from a mostly diagonal matrix ($\zeta = -1$) to one with strong off-diagonal features ($\zeta = 1$) clearly shows how eccentricity affects the structure of the spectral correlation of SGWB. When binaries are nearly circular, the SGWB fluctuations behave like uncorrelated Poisson noise. In contrast, eccentric binaries produce clear correlations between different frequencies, providing a new way to study their orbital properties.

The presence of such off-diagonal spectral correlations in PTA data would provide direct evidence for the eccentric population of binaries contributing to the PTA band. Moreover, the detailed covariance structure encodes information about the eccentricity distribution, enabling constraints on the eccentricity distribution and offering insights into the physical mechanisms driving the dynamical evolution of SMBHBs.

\section{Prospect \lowercase{{\large of}} detecting eccentricity distribution \lowercase{{\large of}} SMBHB population} \label{prosp}

\begin{figure}
  \centering
  \subfigure[]{\label{fig:SNR0}
    \includegraphics[width=0.45\linewidth,trim={0.cm 0 0 0.cm},clip]{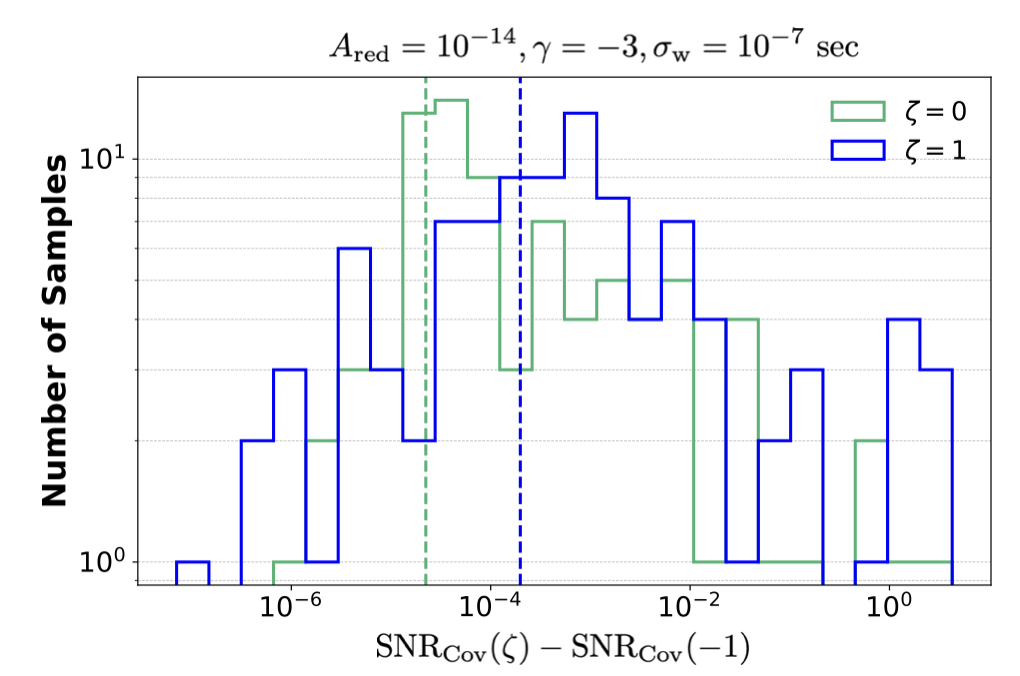}}
  \subfigure[]{\label{fig:SNR1}
    \includegraphics[width=0.45\linewidth,trim={0.cm 0cm 0cm 0.cm},clip]{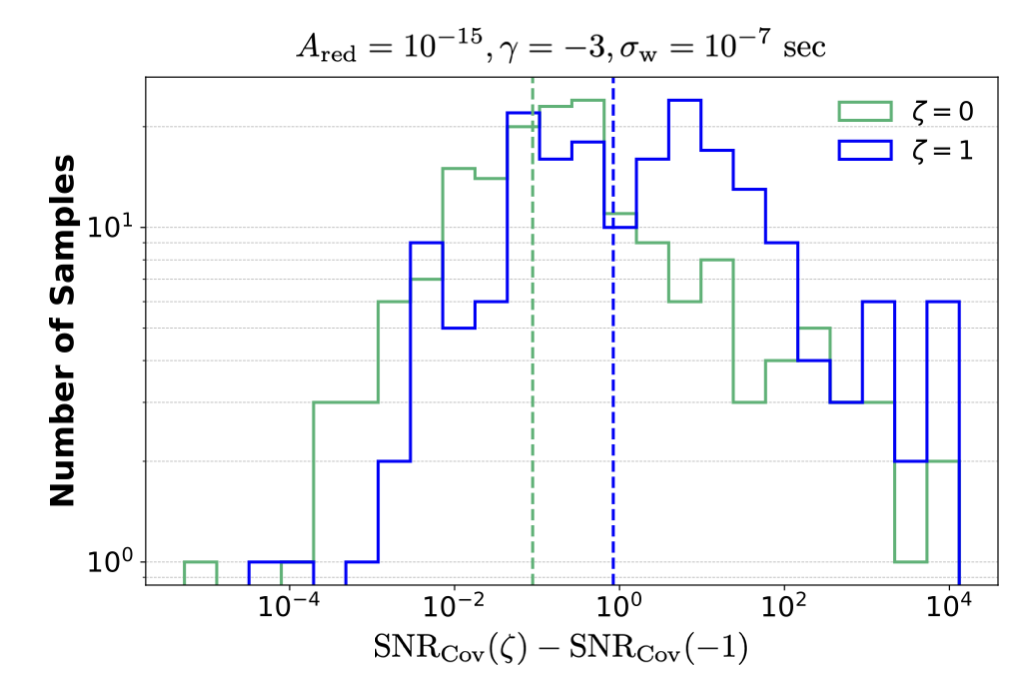}}
  \caption{Histogram showing the distribution of the difference in the SNR of the spectral covariance matrix, defined in Eq.~\ref{SNR_Cov}, relative to the case with $\zeta = -1$. The SNR is computed assuming 2000 isotropically distributed pulsars with identical timing residual noise. Panels (a) and (b) correspond to red noise amplitudes $A_{\rm red} = 10^{-14}$ and $10^{-15}$, respectively, with a fixed spectral index $\gamma = -3$ and white noise level $\sigma_{\rm w} = 10^{-7}$ s. Each histogram is generated using 300 independent Monte Carlo realizations. The dashed vertical line indicates the median value of the distribution.}
  \label{fig:SNR_Cov}
\end{figure}

\begin{figure}
  \centering
  \subfigure[]{\label{fig:SNR0}
    \includegraphics[width=0.45\linewidth,trim={0.cm 0 0 0.cm},clip]{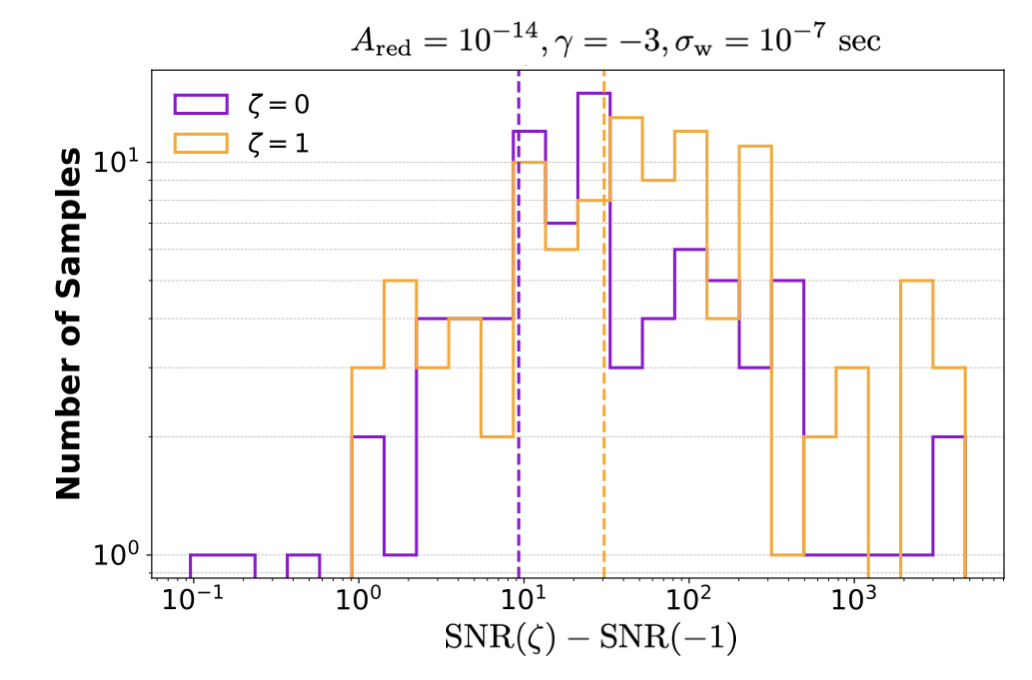}}
  \subfigure[]{\label{fig:SNR1}
    \includegraphics[width=0.45\linewidth,trim={0.cm 0cm 0cm 0.cm},clip]{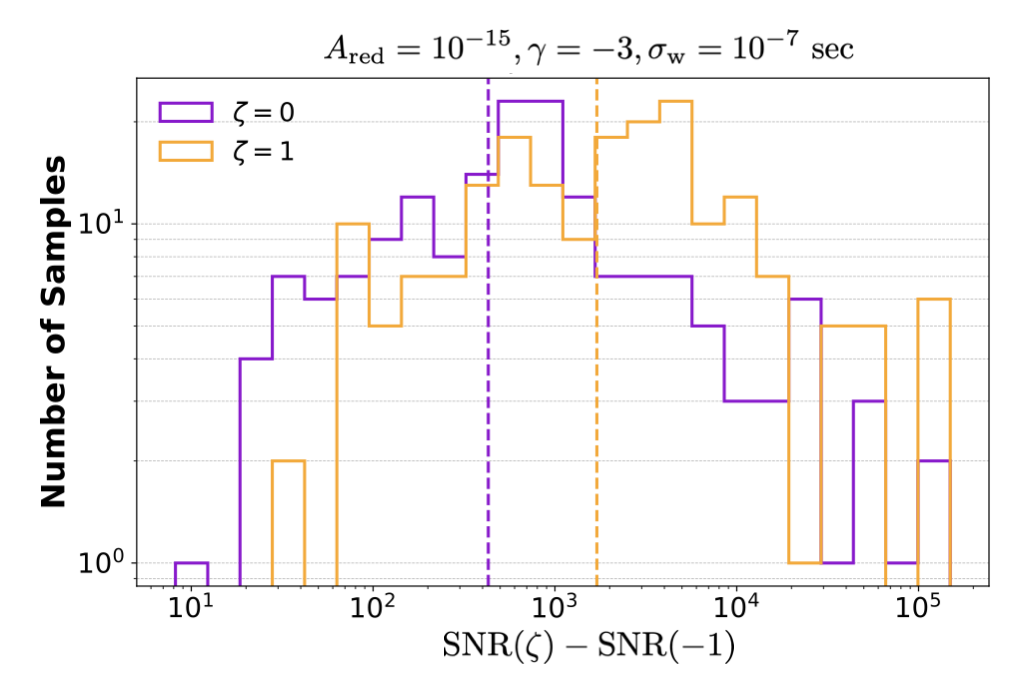}}
  \caption{Histogram showing the distribution of the difference in the combined SNR of the $\Omega_{\rm GW}(f)$ and $C(f_1,f_2)$, defined in Eq.~\ref{SNR_Cov} and Eq.~\ref{SNR_Omega}, relative to the case with $\zeta = -1$. The SNR is computed assuming 2000 isotropically distributed pulsars with identical timing residual noise. Panels (a) and (b) correspond to red noise amplitudes $A_{\rm red} = 10^{-14}$ and $10^{-15}$, respectively, with a fixed spectral index $\gamma = -3$ and white noise level $\sigma_{\rm w} = 10^{-7}$ s. Each histogram is generated using 300 independent Monte Carlo realizations. The dashed vertical line indicates the median value of the distribution.}
  \label{fig:SNR_Omega+Cov}
\end{figure}

To quantify the detectability of eccentricity in the SMBHB population, we compute the signal-to-noise ratio (SNR) of the SGWB signal. The SNR of $\Omega_{\rm GW}(f)$ can be written as

\begin{equation}
    {\rm{SNR}}_{\rm \Omega} = \sqrt{\sum\limits_{f} \frac{\Omega_{\rm GW}^{2}(f)}{\Sigma^2_{\rm \Omega}(f)}},
    \label{SNR_Omega}
\end{equation}
where $\Sigma^2_{\rm \Omega}(f)$ represents the variance in the measurement of $\Omega_{\rm GW}(f)$. $\Sigma^2_{\rm \Omega}(f)$ is computed using the package \texttt{hasasia} \citep{Hazboun2019Hasasia} with the noise model described in Appendix. \ref{sec:timing_noise}. Similarly, we define the  SNR of the spectral covariance matrix, $C(f_1,f_2)$ as
\begin{equation}
    {\rm{SNR}}_{\rm Cov} = \sqrt{\sum\limits_{f_1 \geq f_2} \frac{C^{2}(f_1,f_2)}{\Sigma^2_{\rm cov}(f_1,f_2)}},
    \label{SNR_Cov}
\end{equation}
where $\Sigma^2_{\rm cov}(f_1, f_2)$, derived in Appendix \ref{sec:Pix_noise}, represents the variance in the cross-frequency correlation measurement due to timing residual noise, in the absence of any GW signal. To assess the SNR associated with eccentricity-induced spectral correlations, we compute the difference in SNR between a given eccentricity model and the baseline model with $\zeta = -1$, i.e., $\text{SNR}(\zeta) - \text{SNR}(\zeta = -1)$. This differential SNR is evaluated by keeping the realization of the SMBHB population fixed, with only the eccentricity resampled according to the respective distribution. This quantity isolates the contribution of eccentricity to the spectral covariance, effectively quantifying the additional information content provided by eccentric binary populations relative to the reference case of primarily circular binaries ($\zeta = -1$).

In Fig.~\ref{fig:SNR_Cov}, we show the distribution of this SNR difference for the cases $\zeta = 0$ and $\zeta = 1$, each compared against the $\zeta = -1$, computed over 300 independent Monte Carlo realizations. We take 2000 isotopically distributed pulsars with identical timing residual noise. We explore two noise regimes: (a) a relatively high red noise amplitude $A_{\text{red}} = 10^{-14}$ and (b) a lower red noise amplitude $A_{\text{red}} = 10^{-15}$, with spectral index $\gamma = -3$ and white noise $\sigma_{\rm w} = 10^{-7}$ s. These values are motivated by the noise characteristics observed in current PTA experiments and those anticipated in future observations \citep{agazie2023nanograv_noise,babak2024forecasting}. The analysis is carried out using a HEALPix pixelation scheme with $N_{\rm side} = 16$, corresponding to approximately 2000 independent pulsar baselines. A detailed description of the noise model is provided in Appendix~\ref{sec:timing_noise}.

For the high red noise case ($A_{\text{red}} = 10^{-14}$), distinguishing between eccentricity distributions proves to be challenging. The SNR difference between $\zeta = 0$ and $\zeta = -1$ is small, with a median value (indicated by vertical dashed lines) near zero, indicating that the induced spectral correlations from moderate eccentricity are likely to be buried under noise fluctuations. In contrast, the lower-noise scenario ($A_{\text{red}} = 10^{-15}$) reveals a substantial improvement in detectability. The SNR difference increases significantly, with median values of approximately 0.1 for $\zeta = 0$ and nearly 1 for $\zeta = 1$. This indicates the possibility of detecting the spectral correlation signature induced by eccentricity. When the red noise amplitude is reduced to $A_{\text{red}} \lesssim 10^{-15}$ and the number of well-timed pulsars approaches $\sim 2000$—a scenario expected to be achievable with the Square Kilometre Array (SKA) \citep{smits2009pulsar,maartens2015overview}—PTAs will gain significant statistical power to distinguish between different eccentricity distributions. These results highlight the potential of next-generation PTA experiments to not only detect the nHz SGWB but also to extract rich information about the dynamical properties of its astrophysical sources.

In Fig. \ref{fig:SNR_Omega+Cov}, we show similar results as Fig. \ref{fig:SNR_Cov} but the SNR is now the combined contribution from $\Omega_{\rm GW}(f)$ and $C(f_1,f_2)$ as defined in Eq. \ref{SNR_Omega} and Eq. \ref{SNR_Cov} ($\rm{SNR}^{2} = \rm{SNR}^{2}_{\rm \Omega} + \rm{SNR}^{2}_{\rm Cov}$). While Fig. \ref{fig:SNR_Cov} quantifies the relative enhancement in spectral correlation due to eccentricity, Fig. \ref{fig:SNR_Omega+Cov} captures the absolute detectability of signal by combining both $\Omega_{\rm GW}(f)$ and $C(f_1,f_2)$. 
Notably, the median SNR values shown here  are orders of magnitude larger than those in Fig. \ref{fig:SNR_Cov}. This is because the SNR associated with the total GW energy density spectrum $\Omega_{\mathrm{gw}}(f)$ is intrinsically much larger than the SNR of the cross-frequency $C(f_1,f_2)$, which is a second-order statistic. Even for the high-noise case ($A_{\mathrm{red}} = 10^{-14}$), panel (a) of Fig. \ref{fig:SNR_Omega+Cov} shows that the median SNR can reach of the order of 10 for both $\zeta = 0$ and $\zeta = 1$, implying that the overall signal is detectable even if the eccentricity-dependent structure is not clearly resolvable. In the lower-noise regime shown in panel (b), the median SNR for $\zeta = 1$ reaches values above $10^3$, indicating a strong signal and a significant potential for detecting eccentricity-induced features in the SGWB. Although the SNR associated with $\Omega_{\mathrm{gw}}(f)$ is orders of magnitude higher than that of the spectral covariance $C(f_1, f_2)$, the covariance matrix encodes complementary information that is not captured by the mean spectrum alone. In particular, $C(f_1, f_2)$ is sensitive to the spectral correlations induced by multi-harmonic emission from eccentric binaries. This makes it a powerful tool for breaking degeneracies that exists in $\Omega_{\rm GW}(f)$ among several astrophysical parameters that may otherwise remain unconstrained when using only the isotropic SGWB spectrum.

It is also important to highlight the relatively large spread in the distribution of SNR differences. This spread reflects the inherent cosmic variance associated with different realizations of the SMBHB population. Since we ultimately have access to only a single realization of the Universe, this variance sets a fundamental limit on the confidence with which eccentricity-induced spectral correlations can be inferred from PTA data. Nevertheless, the presence of a statistically significant median offset in the SNR distribution, particularly for highly eccentric populations, suggests that even a single realization—as in our Universe—can carry measurable imprints of the underlying eccentricity distribution. This underscores the importance of incorporating eccentricity as a key parameter in future PTA data analyses.

It is also important to note that the SGWB signal in the nHz regime is often dominated by a small number of exceptionally bright sources, rather than a truly diffuse background. These individually loud binaries, particularly those with high eccentricities and favorable orientations, can inject significant power into multiple frequency bins and substantially shape the spectral structure of the signal. Understanding the role of such dominant sources is therefore crucial because their presence introduces non-Gaussian features in the SGWB and can bias population-level inferences. A detailed investigation of the impact of these bright individual sources on the SGWB statistics, and strategies to identify and characterize them in PTA data, will be pursued in future work.

\section{Conclusion} \label{Conc}

We demonstrated a novel technique for modeling SGWB from a population of eccentric SMBHBs using an accurate numerical relativity–based waveform model, along with observational consequences of orbital eccentricity in SMBHBs on the nHz SGWB. By combining galaxy catalogs from large-scale cosmological simulations with results from the smaller-scale, high-resolution hydrodynamic simulations, we have developed an adaptive multi-scale framework to model the evolution and GW emission of SMBHB populations. In the present paper, we demonstrate this framework using a simple power-law model for the binary separation and eccentricity distribution rather than deriving them from sub-parsec-scale simulations or detailed environmental physics. While this allows broad exploration of eccentric binary populations, it does not capture the full complexity of binary-environment interactions or their correlation with host galaxy properties. In future work, we will extend this framework by incorporating host-galaxy–dependent binary parameters, such as eccentricity and spin, as well as joint distributions of binary separation and eccentricity to better reflect the underlying astrophysics. 

Using the \texttt{MICECAT} simulated galaxy catalog and the SMBH–galaxy relation inferred from the \texttt{ROMULUS25} simulation, combined with an eccentric time-domain waveform model called \esigmahm{}, we constructed SGWB maps and introduced the spectral covariance matrix as a novel observable to characterize the anisotropic SGWB. Unlike analytical waveform models, which often rely on simplifying assumptions and do not account for the spectral features introduced by finite frequency resolution due to the limited observation time of PTA experiments, our approach captures these effects. This leads to a more faithful and realistic representation of the GW signal emitted by SMBHB populations. We demonstrated that a population of eccentric binaries generates strong off-diagonal correlations in the spectral covariance matrix of SGWB density, in contrast to the diagonal structure produced by a population of nearly circular binaries. 
By defining SNR for these correlations, we quantitatively assessed their detectability across different eccentricity distributions and noise scenarios. Our analysis revealed that, although current PTA noise levels limit our ability to detect such correlations, future PTA datasets with improved sensitivity and increased pulsar coverage (as expected from the SKA) can robustly distinguish between different eccentricity populations. We emphasized the significance of the spread in SNR distribution due to cosmic variance. Since we have access to only a single realization of the universe, this intrinsic scatter imposes a fundamental limit on our statistical inference. Nonetheless, the median of the SNR indicates that measurable imprints of eccentricity are likely to be present in PTA data, even from a single universe realization.

This work provides a solid foundation for modeling the nHz SGWB with eccentric binaries, an important step toward accurately interpreting data from PTAs. By capturing both the amplitude and spectral correlation of the SGWB signal, our framework goes beyond traditional isotropic models and enables more detailed characterization of the binary population. In particular, the ability to model spectral correlations opens a new observational for studying the population of SMBHB. This modeling will be especially useful for future nHz GW analysis because it provides a framework for jointly inferring source population properties and breaking degeneracies that affect current SGWB interpretations.

In future work, we plan to improve this framework by adding more detailed models of environmental effects, as well as spin dynamics. These additions will be crucial for making reliable astrophysical conclusions. In summary, our framework marks a major step forward in the study of realistic SGWB from a population of SMBHB. By linking large-scale cosmic structure, galaxy evolution, and accurate waveform physics, it opens new possibilities for uncovering the astrophysical nature of SMBHBs from future nHz GW observations.

\section*{Acknowledgments}
This work is part of the $\langle \texttt{data|theory}\rangle$ \texttt{Universe-Lab} which is supported by the TIFR and the Department of Atomic Energy, Government of India. MS and SM would like to thank the $\langle \texttt{data|theory}\rangle$ \texttt{Universe-Lab} for providing computing resources. AM and PK acknowledge support of the Department of Atomic Energy, Government of India, under project no. RTI4001. PK also acknowledges support by the Ashok and Gita Vaish Early Career Faculty Fellowship at the International Centre for Theoretical Sciences. A part of the computations was performed on the Sonic HPC cluster at ICTS. VS, and AB acknowledge support from the Natural Sciences and Engineering Research Council of Canada (NSERC) through its Discovery Grant program.  AB also acknowledges support from the Infosys Foundation via an endowed Infosys Visiting Chair Professorship at the Indian Institute of Science.  AB, TQ, and MT gratefully acknowledge partial support via NSF award AST-1514868. CKM acknowledges the support of SERB’s Core Research Grant No. CRG/2022/007959. This work has made use of CosmoHub \citep{2017ehep.confE.488C, TALLADA2020100391}, developed by PIC (maintained by IFAE and CIEMAT) in collaboration with ICE-CSIC. It received funding from the Spanish government (grant EQC2021-007479-P funded by MCIN/AEI/10.13039/501100011033), the EU NextGeneration/PRTR (PRTR-C17.I1), and the Generalitat de Catalunya. The Romulus simulation is part of the Blue Waters sustained-petascale computing project, which is supported by the National Science Foundation (via awards OCI-0725070, ACI-1238993, and OAC-1613674) and the state of Illinois. Blue Waters is a joint effort of the University of Illinois at Urbana-Champaign and its National Center for Supercomputing Applications. Resources supporting this work were also provided by the (a) NASA High-End Computing (HEC) Program through the NASA Advanced Supercomputing (NAS) Division at Ames Research Center; (b) Extreme Science and Engineering Discovery Environment (XSEDE), supported by National Science Foundation grant number ACI-1548562; and (c) Digital Research Alliance of Canada (alliancecan.ca). The authors would also like to acknowledge the use of the following Python packages in this work: Numpy \citep{van2011numpy,2020NumPy-Array}, Scipy \citep{jones2001scipy,virtanen2020scipy}, Matplotlib \citep{hunter2007matplotlib}, Astropy \citep{robitaille2013astropy,price2018astropy,2022ApJ...935..167A}, Hasasia \citep{Hazboun2019Hasasia}, and Ray \citep{moritz2018ray}.

\bibliographystyle{aasjournal}
\bibliography{main}

\appendix
\section{Results \lowercase{{\large from}} \lowercase{{\large a}} Model \lowercase{{\large with}} Evolutionary Coupling \lowercase{{\large between}} Eccentricity \lowercase{{\large and}} Binary Separation} \label{sec:ecc_evol}

In the main analysis presented in this paper, we assume that the binary separation and the corresponding orbital eccentricity are independent of each other. Specifically, we do not account for the circularization of binary orbits over time. Instead, the semi-major axis and eccentricity of each binary are independently drawn from power-law distributions. We motivate this assumption by recalling that the evolution of eccentricity is also influenced by environmental effects in addition to GW-driven hardening \citep{chen2017efficient,gualandris2022eccentricity,siwek2024signatures}.

In the case of purely GW-driven evolution, the binary orbits are expected to circularize.  In this section, we present results from an astrophysical scenario in which the semi-major axis (or, equivalently, the orbital frequency) and eccentricity of each binary are determined self-consistently by evolving the orbit purely under GW radiation. This approach captures the natural coupling between orbital separation and eccentricity that arises during binary evolution.

Each binary is initialized with a fixed eccentricity at a given orbital frequency and is subsequently evolved solely through GW-driven orbital decay. We then investigate how this physically motivated evolution impacts key observables discussed throughout this paper, namely the SGWB energy density spectrum, $\Omega_{\rm GW}(f)$, and the structure of the spectral covariance matrix, $C_{\rm N}(f_1, f_2)$. The evolution of the binary orbit and eccentricity, purely by GW radiation, is given by \citep{peters1963gravitational, peters1964gravitational} 

\begin{equation}
    \frac{d f_{\rm orb}}{dt_r} = \frac{96}{5 c^{5}} (2\pi)^{8/3} G^{5/3} \mathcal{M}_{c}^{5/3} f_{\rm orb}^{11/3} ~\mathbb{F}(e),
\end{equation}

\begin{equation}
    \frac{d e}{dt_r} = - \frac{1}{15 c^5} (2\pi)^{8/3} G^{5/3} \mathcal{M}_{c}^{5/3} f_{\rm orb}^{8/3}~ \mathbb{G}(e),
\end{equation}

where
\begin{equation}
    \mathbb{F}(e) = \frac{1 + \frac{73}{24} e^2 + \frac{37}{96} e^4}{(1 - e^2)^{7/2}}
\end{equation}

\begin{equation}
    \mathbb{G}(e) = \frac{304 e + 121 e^3}{(1 - e^2)^{5/2}},
\end{equation}
where $f_{\rm orb}$ is the orbital frequency, $t_{r}$ is the time in the source’s cosmic rest frame, and $\mathcal{M}_{c}$ is the chirp mass. To compute the number density of sources orbiting at a given orbital frequency $f_{\rm orb} $, and corresponding eccentricity, we solve the above coupled differential equations for $ \frac{dt_{\rm r}}{df_{\rm orb}}(f_{\rm orb}) $ and $ e(f_{\rm orb}) $. The number of sources per unit frequency is assumed to be proportional to the residence time ($\frac{dN}{df_{\rm orb}}(f_{\rm orb}) \propto \frac{dt_{\rm r}}{df_{\rm orb}}(f_{\rm orb})$) at that frequency.

\begin{figure}[ht!]
    \centering
    \includegraphics[width=14cm]{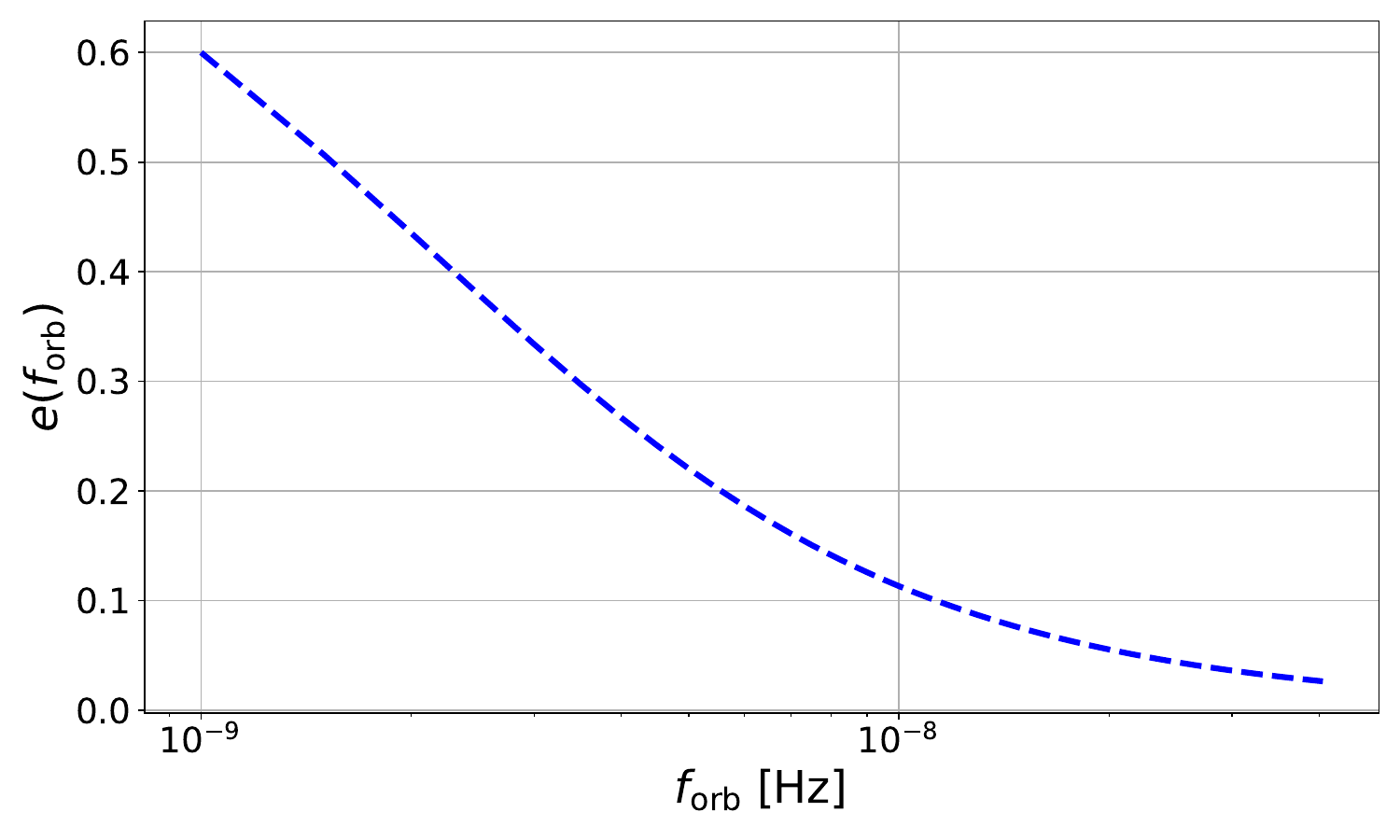}
    \caption{Eccentricity ($e$) of the SMBH binary as a function of its orbital frequency ($f_{\rm orb}$). The binary begins with an initial eccentricity of $e = 0.6$ at $f_{\rm orb} = 10^{-9}$ Hz and evolves solely under the influence of GW radiation.} 
    \label{fig:App:e_vs_f}
\end{figure}

\begin{figure}[ht!]
    \centering
    \includegraphics[width=14cm]{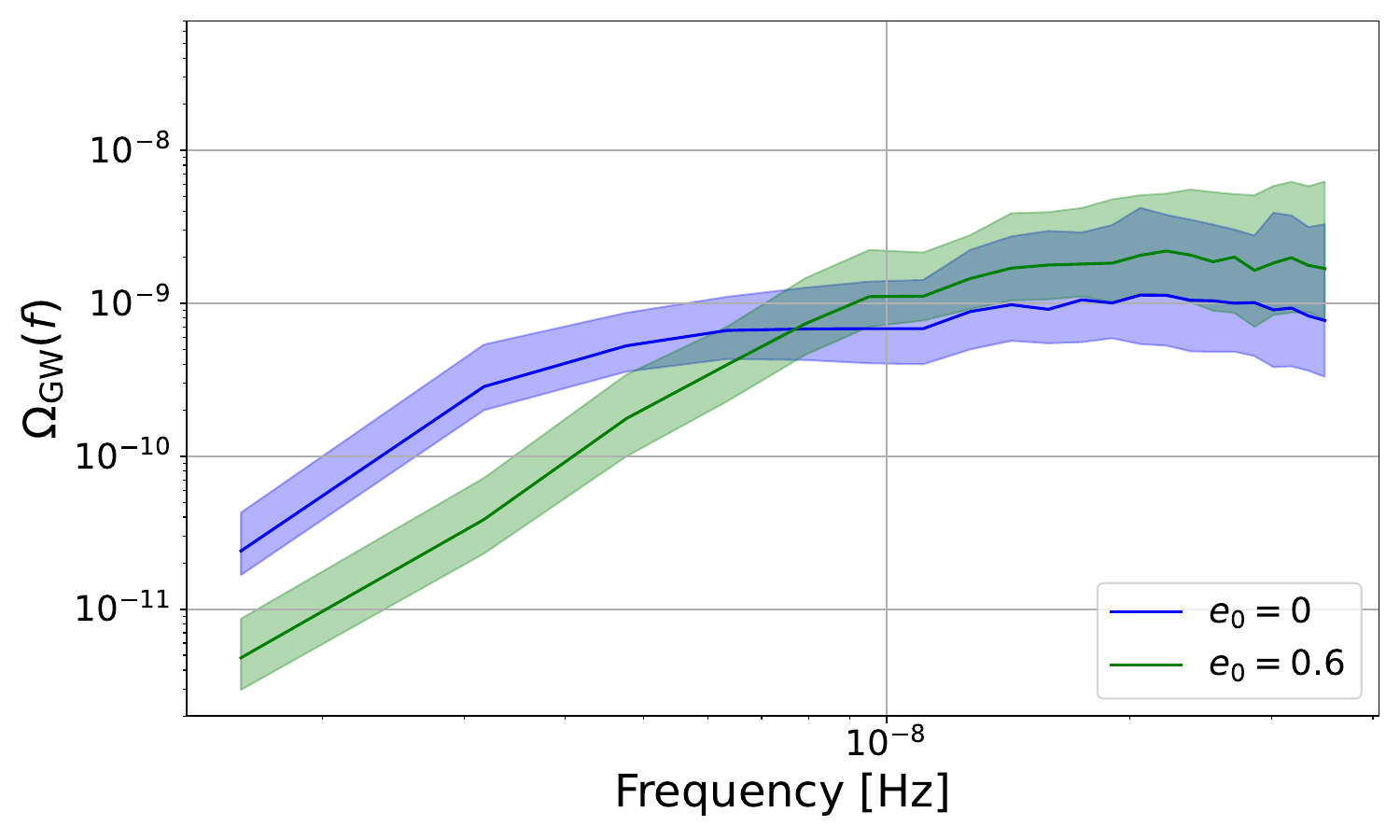}
    \caption{SGWB spectrum, $\Omega_{\rm GW}(f)$, for a population of binaries with (a) circular orbit ($e_{0} = 0$) (b) initial eccentricity, $e_{0} = 0.6$ at $f_{\rm orb} = 10^{-9}$ Hz. The solid lines represent the median $\Omega_{\rm GW}(f)$ obtained from 300 Monte Carlo realizations, while the shaded regions denote the $68\%$ credible intervals.}
    \label{fig:App:Omega}
\end{figure}

\begin{figure}
  \centering
  \subfigure[]{\label{fig:CovMinus1}
    \includegraphics[width=0.45\linewidth,trim={0.cm 0 0 0.cm},clip]{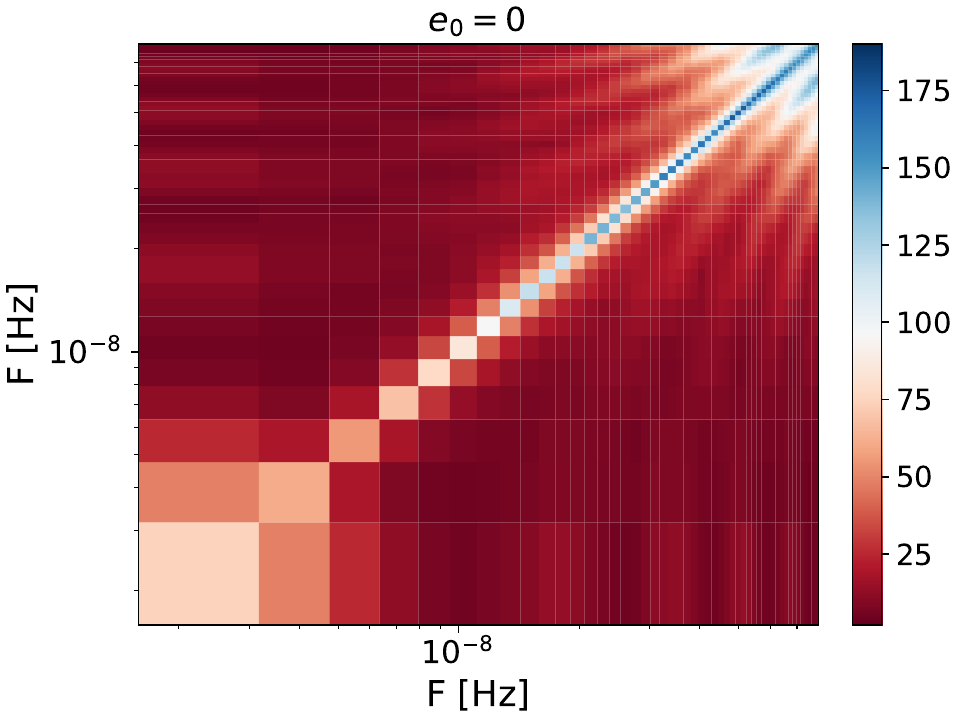}}
  \subfigure[]{\label{fig:Cov0}
    \includegraphics[width=0.45\linewidth,trim={0.cm 0cm 0cm 0.cm},clip]{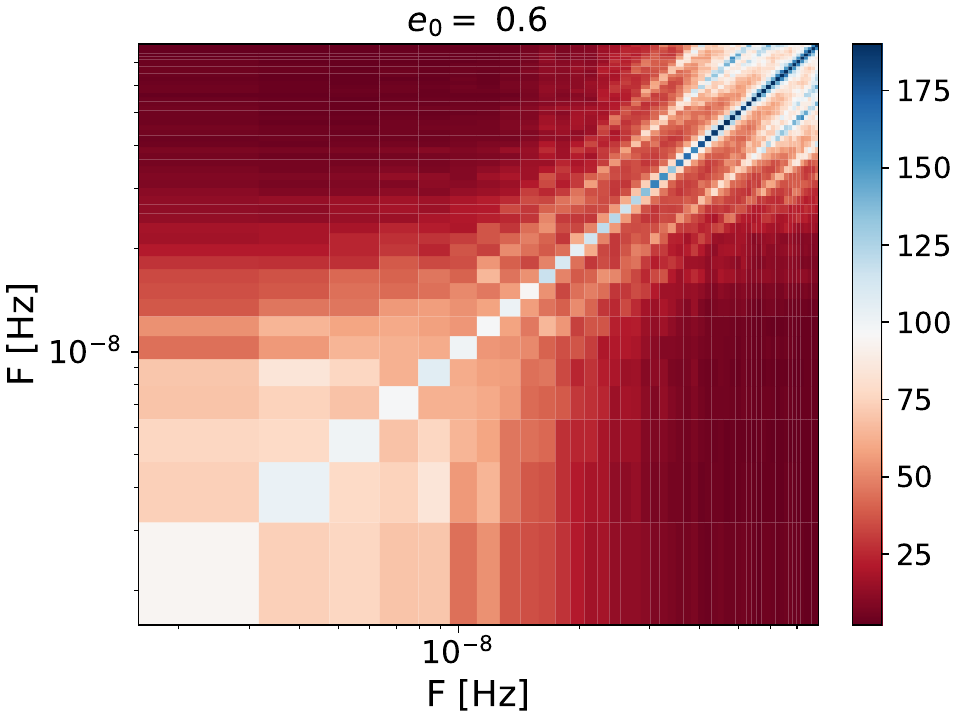}}
  \caption{Normalized spectral covariance matrix $C_{\rm N}(f_1, f_2)$ (defined in Eq. \ref{norm_cross}) for a population of binaries with (a) circular orbit ($e_{0} = 0$) (b) initial eccentricity, $e_{0} = 0.6$ at $f_{\rm orb} = 10^{-9}$ Hz.}
  \label{fig:App:Matrix}
\end{figure}

We assume a fiducial population with $ e = 0 $ (i.e., circular binaries) that contains a total of 25000 sources in the PTA band. These sources are distributed according to the residence time, i.e., proportional to $\frac{dt_{\rm r}}{df_{\rm orb}}(f_{\rm orb})$ and then eccentricity is assigned to each binary using the relation, $e(f_{\rm orb})$. For populations with a non-zero initial eccentricity $ e_0 $ defined at a reference orbital frequency $f_{\rm orb}$, the number of sources at each frequency is computed by scaling the circular population using the ratio of residence times.

\begin{equation}
\frac{dN}{df_{\rm orb}}(f_{\rm orb} \mid e_0 \neq 0) = \Bigg(\frac{dt_{\rm r}}{df_{\rm orb}}(f_{\rm orb} \mid e_0 \neq 0) \left/ \frac{dt_{\rm r}}{df_{\rm orb}}(f_{\rm orb} \mid e_0 = 0) \right.\Bigg) \times  \frac{dN}{df_{\rm orb}}(f_{\rm orb} \mid e_0 = 0).
\label{eq:rel_reside_time}
\end{equation}
Since we are interested in comparing the results with respect to the circular orbit case, we have modeled the number of sources in the eccentric population as described above. This approach allows us to capture how the distribution of source frequencies is modified due to the faster orbital decay associated with eccentricity, for the same initial population as for the circular case.

In Fig. \ref{fig:App:e_vs_f}, we show the eccentricity $e$ as a function $f_{\rm orb}$. The binary starts with the eccentricity of 0.6 at $f_{\rm orb} = 10^{-9}$ Hz and evolves under GW radiation. As the orbital frequency increases to $10^{-8}$ Hz, the eccentricity decreases to approximately 0.1, reflecting the gradual circularization of the binary during its inspiral. In Fig. \ref{fig:App:Omega}, we compare the energy density spectrum $\Omega_{\rm GW}(f)$ for two populations of binaries: one composed of circular binaries ($e_0 = 0$) and the other initialized with an eccentricity of $e_0 = 0.6$ at an orbital frequency of $f_{\rm orb} = 10^{-9}$ Hz. The solid curves represent the median spectrum over 300 realizations, while the shaded bands indicate the $68\%$ credible intervals. At lower frequencies, the $\Omega_{\rm GW}(f)$ for eccentric binaries (green) lies below that of the circular population (blue). This is because eccentric binaries evolve more rapidly due to enhanced GW radiation, spending less time at larger separations where they emit at lower frequencies. Consequently, fewer eccentric binaries contribute to the low-frequency part of the spectrum.

As binaries inspiral and gradually circularize, the relative residence time between eccentric and circular binaries increases and asymptotically approaches unity. This leads to a convergence in the population sizes of the eccentric and circular cases. Although the number of eccentric binaries remains slightly smaller than circular binaries at higher frequencies, the larger GW luminosity associated with eccentric orbits compensates for this, resulting in a higher $\Omega_{\rm GW}(f)$ above $\sim 10^{-8}$ Hz. This feature is absent when eccentricity and orbital frequency are drawn independently from power-law distributions, as shown in Fig. \ref{fig:Omega}. In that case, the shape of the spectrum remains largely unchanged regardless of the eccentricity distribution, due to the lack of coupling between eccentricity and frequency in the source population. Furthermore, the flattening at higher frequencies observed in the eccentricity-orbital separation coupled scenarios arises because the number of binaries emitting dominantly at those frequencies becomes significantly smaller than in the power-law distribution cases discussed in the main analysis. As a result, the distribution of $\Omega_{\rm GW}(f)$ at high frequencies becomes increasingly skewed toward larger values, causing the median to lie substantially below the mean in that range.

Fig. \ref{fig:App:Matrix} shows the normalized spectral covariance matrix $C_{\rm N}(f_1, f_2)$ for the two representative cases, $e_0 = 0$ and $e_0 = 0.6$. The spectral correlations are more pronounced for the eccentric population ($e_0 = 0.6$), particularly at lower frequencies. This is because eccentric binaries emit over a broader range of harmonics, naturally introducing spectral correlations. As the binaries inspiral and circularize, these correlations diminish, leading to a more diagonal structure at higher frequencies. For the circular case ($e_0 = 0$), spectral correlations are still present, although weaker, arising primarily due to the finite frequency resolution of PTA observations, as discussed in Sec.\ref{Waveform}.

In contrast to the results shown in the main analysis of the paper, where the covariance matrix structure remains broadly uniform, the eccentricity–binary separation coupled scenario introduces a frequency-dependent evolution in the matrix structure. In particular, the off-diagonal correlations become less pronounced at higher frequencies due to the circularization of the binaries.

We emphasize, however, that such circularization may not represent the full astrophysical picture. In realistic environments, interactions of the binary with surrounding stars, gas, or circumbinary disks can act to increase its eccentricity, as suggested by \cite{chen2017efficient, gualandris2022eccentricity, siwek2024signatures}. Thus, while our example illustrates the impact of GW radiation alone, more comprehensive modeling would need to incorporate environmental effects.

\section{Timing Residual Noise Model}
\label{sec:timing_noise}

The total noise power spectrum of pulsar timing residuals can be modeled as the sum of white and red noise contributions:
\begin{equation}
    P(f) = P_{\rm w}(f) + P_{\rm r}(f),
\end{equation}
where $P_{\rm w}(f)$ represents the white noise component and $P_{\rm r}(f)$ captures the red noise contributions. The white noise term is given by
\begin{equation}
    P_{\rm w}(f) = 2 \, \Delta t \, \sigma_{\rm w}^2,
\end{equation}
where $\Delta t$ denotes the cadence of pulsar time-of-arrival (TOA) measurements, and $\sigma_{\rm w}$ is the TOA uncertainty. This term models uncorrelated, time-independent measurement noise.

The red noise component, $P_{\rm r}(f)$, encompasses both intrinsic pulsar noise and the contribution from GW
\begin{equation}
    P_{\rm r}(f) = P_{\rm I}(f) + P_{\rm gw}(f),
\end{equation}
where $P_{\rm I}(f)$ models the intrinsic red noise, and $P_{\rm gw}(f)$ represents the GW-induced noise.

The intrinsic red noise is typically modeled as a power-law:
\begin{equation}
    P_{\rm I}(f) = \frac{A_{\rm red}^2}{12\pi^2} \left(\frac{f}{f_{\rm 1yr}}\right)^{\gamma} f_{\rm 1yr}^{-3},
\end{equation}
where $A_{\rm red}$ is the red noise amplitude, $\gamma$ is the spectral index, and $f_{\rm 1yr} = 1/\text{yr}$ is a reference frequency corresponding to a period of one year.

$P_{\rm gw}(f)$ can be expressed as
\begin{equation}
    P_{\rm gw}(f) = \frac{h_c^2(f)}{12 \pi^2} f^{-3},
\end{equation}
where $h_c(f)$ is the characteristic strain of the GW background.

\section{Derivation \lowercase{{\large of}} \lowercase{{\large the}} pixel level noise}
\label{sec:Pix_noise}

The cross-correlation of the Fourier transform of the timing residuals from two pulsars I and J is given by:

\begin{align}
    \langle \tilde{r}_I(f) \tilde{r}_J^*(f) \rangle \propto  & \int d\hat{n} ~ \Omega_{\text{gw}}(f,\hat{n}) \\ & \nonumber \Big(\mathcal{F}^{+}_{I}(\hat{n}) \mathcal{F}^{+}_{J}(\hat{n}) + \mathcal{F}^{\times}_{I}(\hat{n}) \mathcal{F}^{\times}_{J}(\hat{n})\Big),
\end{align}
where  
\begin{equation}
     \mathcal{F}^{A}_{I}(\hat{n}) = \frac{1}{2} \frac{\hat{p_I}^{a}\hat{p_I}^{b}}{1+\hat{n}  \cdot \hat{p_I}} e^{A}_{ab}.
\end{equation}

The overlap reduction function between pulsars I and J, $\Gamma_{IJ}$, in an equal-pixel basis, called radiometer basis \citep{mitra2008gravitational,thrane2009probing} can be expressed as

\begin{equation}
    \Gamma_{IJ}(f) = \sum\limits_{k} ~\Omega_{\rm GW}(f,\hat{n}_k) \Big(\mathcal{F}^{+}_{I}(\hat{n}_k) \mathcal{F}^{+}_{J}(\hat{n}_k) + \mathcal{F}^{\times}_{I}(\hat{n}_k) \mathcal{F}^{\times}_{J}(\hat{n}_k)\Big).
\end{equation}
We can represent this in a matrix form as 
\begin{equation}
    \mathbf{\Gamma}_{IJ} = \textbf{R} ~ \mathbf{\Omega},
\end{equation}
where $\textbf{R}^{IJ}_{k}  \equiv  \sum\limits_A \mathcal{F}^{A}_{I}(\hat{n}_k)  \mathcal{F}^{A}_{J}(\hat{n}_k) $ and $\mathbf{\Omega}_{k} \equiv \Omega_{\rm GW}(f,\hat{n}_k)$. We can write the likelihood function for the cross-correlations as

\begin{equation}
    \mathcal{P}(\hat{\textbf{X}}|\mathbf{\Omega}) \propto \textbf{exp}[\frac{-1}{2} (\hat{\textbf{X}}(f) - \textbf{R} \mathbf{\Omega}(f))^{T} \Sigma_{\rm cc}^{-1}(f) (\hat{\textbf{X}}(f) - \textbf{R} \mathbf{\Omega}(f))],
\end{equation}
where $\hat{\textbf{X}}_{IJ}(f) \propto \tilde{r}_I(f) \tilde{r}_J^*(f)$, and $\Sigma_{\rm cc}(f)$ is the cross-correlation covariance matrix.

We denote the uncertainty in the measurement of $\Omega_{\rm GW}(f,\hat{n}_k)$ by $\mathbf{\Sigma}_{\rm pix}(f, \hat{n}_k)$. The uncertainty in the measurement of the spectral cross-correlation $C(f_1,f_2)$ (defined in Eq. \ref{cross}), obtained by optimally combining the cross-correlation signals from all sky pixels using an inverse-variance weighted mean, is given by

\begin{equation}
    \Sigma^{2}_{\rm cov}(f_1,f_2) =  \frac{1}{\sum\limits^{N_{\rm pix}}_{k}\frac{1}{\sum_{\rm pix}(f_1,\hat{n}_k) \times\sum_{\rm pix}(f_2,\hat{n}_k)} },
\end{equation}
where $N_{\rm pix}$ denotes the number of pixels. For isotropic noise, $\Sigma^{2}_{\rm cov}$ can be written as

\begin{equation}
    \Sigma^{2}_{\rm cov}(f_1,f_2) = \frac{\sum_{\rm pix}(f_1) \times \sum_{\rm pix}(f_2)}{N_{\rm pix} } 
    \label{cov_noise},
\end{equation}

\section{Accuracy estimates \lowercase{{\large of}} \inspiralesigma{} waveform model}
\label{appendix:accuracy_esigmahm}
In this work, we have used the eccentric waveform model \inspiralesigma{} \citep{paul2024esigmahm} to simulate the GW waveforms from SMBHBs with starting eccentricities going util $e=0.8$. As mentioned in Sec. \ref{Waveform}, \inspiralesigma{} contains eccentric spinning and non-spinning corrections up to 3PN order \citep{Henry:2023tka} in the radiative dynamics of the model. These corrections include the instantaneous contributions as well as the hereditary contributions. However, unlike the instantaneous contributions which are exact in eccentricity, the hereditary contributions are expanded in eccentricity and then resummed. Although it has been shown that these resummed expressions perform better than their bare eccentricity expanded versions even for eccentricities close to 1 (see Figure 1 of \cite{Henry:2023tka}), we perform an accuracy analysis of \inspiralesigma{} here to have some quantitative estimates to gauge the accuracy of the model in the high eccentricity regime. 

\begin{figure}
    \centering
    \includegraphics[width=14cm]{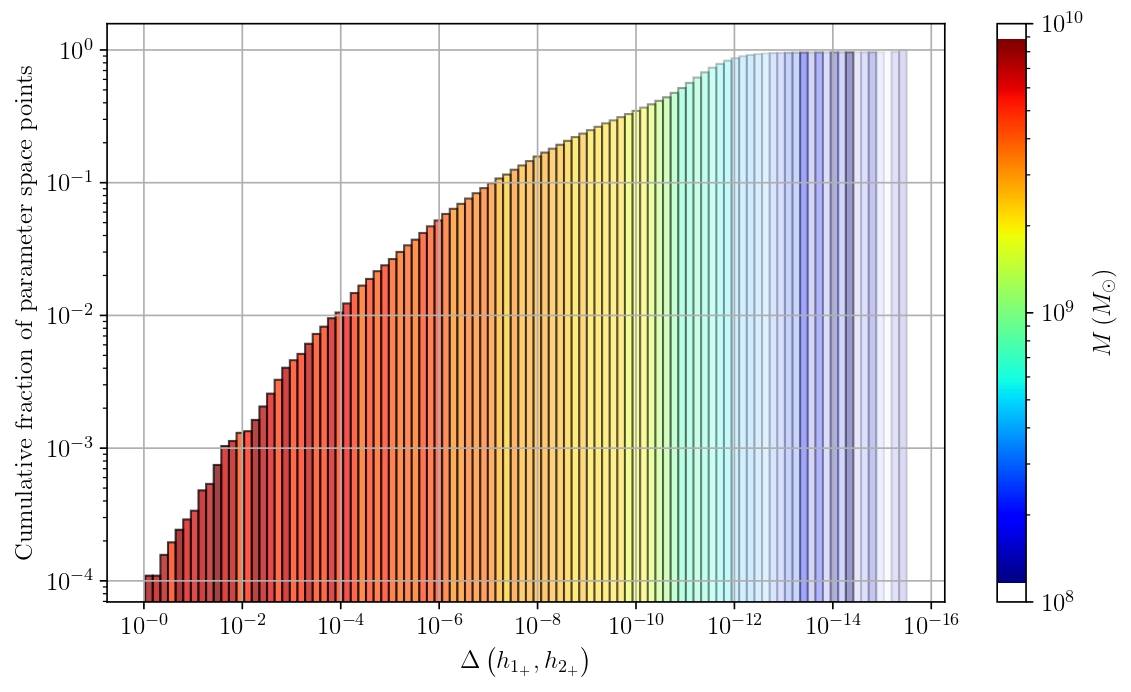}
    \caption{A histogram of the cumulative fraction of parameter space points as one accumulates points with increasingly lower relative differences (c.f. Eq. \eqref{eq:relative-difference}) between the $+$ polarizations obtained from \inspiralesigma{} and the ``reduced" version of \inspiralesigma{} from which all the eccentric 3PN contributions from the radiative dynamics have been removed. The histogram bins are colored according to the median value of total mass ($M$) for the parameter space points corresponding to that bin. Further, the opacity of each bin signifies the median eccentricity corresponding to that bin, with a higher opacity signifying a higher median eccentricity.}
    \label{fig:ESIGMA-accuracy}
\end{figure}

 We compute the relative difference between the $+$ GW polarizations obtained from \inspiralesigma{} (denoted by $h_{1_+}$) and the version of \inspiralesigma{} from which we have removed all the eccentric 3PN contributions from the radiative dynamics (denoted by $h_{2_+}$). First, we align the waveforms in phase and time by maximizing the overlap between the waveforms, called \textit{mismatch},

\begin{equation}
    \mathcal{M}(h_{1_+}, h_{2_+}) =\max \limits_{\phi_0, t_0 } \frac{\left( h_{1_+}|h_{2_+} e^{i\left( \phi_0 + 2 \pi f t_0 \right)} \right)}{\sqrt{\left( h_{1_+}|h_{1_+} \right)\left( h_{2_+}|h_{2_+} \right)}},
\end{equation}
where we define,
\begin{equation}
    \left(a|b\right) \equiv 4\, \Re \left[ \int_{0}^{f_{\rm{max}}} \tilde{a}(f) \tilde{b}^*(f) \rm{d}f \right]\,,
\label{eq:overlap}
\end{equation}
where $\tilde{a}(f)$ denotes the Fourier transform of a time-domain function $a(t)$, $*$ denotes complex conjugation, $f_{\rm{max}} = f_{\rm{Ny}} \simeq 1/2\Delta t$ is the Nyquist frequency of the waveforms where $\Delta t = 30$ days and the integral is calculated over a frequency grid with spacing $\Delta f=1/T_{\rm{obs}}$ with $T_{\rm{obs}} = 20$ yr\footnote{The time-domain waveforms are zero-padded by appropriate amounts to bring the frequency domain signal over this grid.}. After aligning the two waveforms by applying the relative phase-shift $\phi_0$ and time-shift $t_0$, we compute the relative difference between the two waveforms
\begin{equation}
\Delta(h_{1_+}, h_{2_+}) =  \dfrac{||h_{1_+} - h_{2_+}||}{||h_{1_+}||}, \text{ where}\, ||h|| = \sqrt{\int\limits_{f_{\rm{min}}}^{f_{\rm{max}}} |h(f)|^2 \, df} \, ,  
\label{eq:relative-difference}
\end{equation}
where $f_{\rm{min}} = \Delta f$ and $f_{\rm{max}} = f_{\rm{Ny}} \simeq 1/2\Delta t$.

We compute the GW polarizations from the two waveform models over a parameter space defined by the primary mass $m_1 \in [10^8 M_\odot, 10^{10} M_\odot]$, mass ratio $q = m_2/m_1 \in [0.01, 1]$, eccentricity $e \in [0, 0.8]$, mean anomaly $l \in [0, 2\pi]$, redshift $z \in [0.07, 1.4]$, and inclination angle $i \in [0, \pi]$. The parameters $m_1$ and $q$ are sampled from a uniform distribution in logarithmic space, while $e$, $l$, and $z$ are drawn from uniform distributions. The starting frequency of the waveform, $f_{\rm{start}}$ (defined as twice the orbital frequency), is sampled from a power-law distribution $P(f_{\rm{start}}) \propto f_{\rm{start}}^{-2.6}$. This corresponds to a semi-major axis distribution $P(\mathtt{a} \mid M_*, z) \propto \mathtt{a}^{3.8}$, which is used to generate the source population in this study. The values of $f_{\rm{start}}$ are restricted to the range $[1/T_{\rm{obs}},\, 3 \times 10^{-8}\,\mathrm{Hz}]$.

However, some waveforms become too short at the higher end of the eccentricity range, starting frequency, and total masses, and also exhibit some numerically unstable growth in waveform amplitudes. To weed out such unphysical waveforms from the analysis, we calculate the ratio of the maximum amplitude and the median amplitude for each waveform, i.e.

\begin{equation}
    r = \frac{\max_t A(t)}{\text{median}_t A(t)},
\end{equation}
where $A(t) = \sqrt{h_+^2(t) + h_\times^2(t)}$ is the amplitude of the time-domain waveform. We use this ratio as a discriminator for removing unphysical waveforms by discarding those parameter space points where it becomes higher than a threshold value for either \inspiralesigma{} or the “reduced” version of \inspiralesigma{}. We find that on keeping only those parameter space points which have $r<7$ for both the waveform models yields relative differences $\Delta(h_{1_+}, h_{2_+})$ of less than $1$ (i.e. $100 \%$) across that parameter space between the two models. Hence, we choose to work only with the subset of parameter space where $r<7$ for both the waveform models. We find that this threshold on $r$ discards $\sim 10\%$ of the parameter space points corresponding to high eccentricity values, starting frequency and total masses, where the waveforms become very short. 

We now discuss the variation of the relative difference between the waveforms in this restricted parameter space.

In Fig. \ref{fig:ESIGMA-accuracy}, we show the histogram of the cumulative fraction of parameter space points as one accumulates points with increasingly lower relative differences (c.f. Eq. \eqref{eq:relative-difference}) between the $+$ polarizations obtained from \inspiralesigma{} and the ``reduced" version of \inspiralesigma{}. We observe that only $0.1 \%$ points in the parameter space show relative differences of greater than $1\%$ between the two waveform models. We also show the median total mass value ($M = m_1+m_2$) and the median eccentricity value within each histogram bin via color and its opacity respectively, with higher opacity signifying higher median eccentricity for the parameter space points in that particular bin. As mentioned earlier, we observe that the high $\Delta(h_{1_+}, h_{2_+})$ values correspond to high total mass and high eccentricity sector of the parameter space.  However, since the \texttt{MICECAT} catalog used to generate the population contains very few galaxies with stellar mass $M_* > 10^{12} \, M_{\odot}$, such sources account for less than $0.1\%$ of the total. We find that their inclusion or exclusion has a negligible impact on the resulting SGWB. Therefore, we can safely adopt a maximum eccentricity of $e = 0.8$.

However, we remark that this analysis only compares ESIGMA with its version which has the highest order eccentric contributions removed from its radiative dynamics equations. Given that PN expansions are asymptotic in nature, such an analysis may not reflect the true accuracy of the model, which can only be found out reliably by comparing the model with numerical relativity simulations. Furthermore, \esigmahm{} is in active development to make it more faithful to the existing numerical relativity simulations.

\end{document}